\documentclass[aip,reprint]{revtex4-1}
\usepackage{graphicx}% Include figure files
\usepackage{dcolumn}% Align table columns on decimal point
\usepackage{bm}% bold math
\usepackage{amssymb,amsmath}
\usepackage[utf8]{inputenc}
\usepackage[T1]{fontenc}
\usepackage{mathptmx}
\usepackage{etoolbox}
\usepackage{color}

\usepackage{array,multirow,graphicx}
\usepackage{float}
\usepackage{blindtext}
\usepackage{subfiles}

\usepackage{array}
\usepackage{makecell}
\draft % marks overfull lines with a black rule on the right

\begin{document}

%\preprint{}

\title[Exploring the robustness of permutation entropy analysis]{Exploring the robustness of permutation
entropy analysis to differentiate between
closed-eyes and open-eyes resting states}
% Force line breaks with \\
\author{Juan Gancio}
 %\homepage{http://www.Second.institution.edu/~Charlie.Author.}
\email{juan.gancio@upc.edu}
\affiliation{%
Universitat Politècnica de Catalunya, Departament de Física, Rambla Sant Nebridi 22, Terrassa 08222, Barcelona, Spain.%\\This line break forced% with \\
}%
\author{Natalia López López}
 %\homepage{http://www.Second.institution.edu/~Charlie.Author.}
\affiliation{%
Universitat Politècnica de Catalunya, Departament de Física, Rambla Sant Nebridi 22, Terrassa 08222, Barcelona, Spain.%\\This line break forced% with \\
}
\author{Antonio J. Pons}
 %\homepage{http://www.Second.institution.edu/~Charlie.Author.}
\affiliation{%
Universitat Politècnica de Catalunya, Departament de Física, Rambla Sant Nebridi 22, Terrassa 08222, Barcelona, Spain.%\\This line break forced% with \\
}
\author{Giulio Tirabassi}%
\affiliation{%
Universitat de Girona, Departament de Informàtica, Matemàtica Aplicada i Estadística, Universitat de Girona, Carrer de la Universitat de Girona 6,  Girona 17003, Spain.%\\This line break forced% with \\
}
\author{Cristina Masoller}
 %\homepage{http://www.Second.institution.edu/~Charlie.Author.}
\affiliation{%
Universitat Politècnica de Catalunya, Departament de Física, Rambla Sant Nebridi 22, Terrassa 08222, Barcelona, Spain.%\\This line break forced% with \\
}

\date{\today}% It is always \today, today,

\begin{abstract}
{Electroencephalography (EEG) is a noninvasive technology that is widely used to monitor brain states, and many efforts are focused on developing reliable and efficient data analysis methods for EEG recordings. Here, we apply ordinal analysis to the EEG recording of the resting state of 109 healthy subjects {measured in two different conditions: with eyes closed (EC state) or eyes open (EO state)}. We study the robustness of the temporal permutation entropy ($PE$) and the spatial permutation entropy ($SPE$) with respect to the presence of blinking artifacts in the EO recordings, the duration of the recordings, and the number of EEG channels analyzed. {We perform a paired statistical test to assess whether these quantities differ significantly in the EO and EC recordings of the same subject}. We find that $PE$ and $SPE$ perform surprisingly well, as they detect significant differences when calculated from the raw signals, even within a short time interval ($PE$) or from a reduced number of electrodes ($SPE$).}
\end{abstract}


\maketitle %\maketitle must follow title, authors, abstract and \pacs

\begin{quotation}
{Ordinal analysis is a symbolic method that has been applied to analyze complex, high dimensional data in many different areas (images, art works, textures, cardiac signals, climatological data, financial data, etc.). We have previously applied this method to electroencephalography (EEG) signals of healthy subjects recorded in the resting state, with eyes closed (EC state) or eyes open (EO state). 
Here, we demonstrate that the ordinal method is capable of detecting significant differences in individual subject records when applied to raw measurements (without the need to remove artifacts) covering a short time interval, or from the analysis of data values from a small number of synchronously recorded electrodes.}
\end{quotation}
\section{Introduction}
Since the German psychiatrist Hans Berger invented electroencephalography (EEG), his technique for recording the brain's electrical activity through electrodes placed on the scalp has developed enormously %. The physical origin of EEG-based 
%techniques are very well understood~\citep{Nunez2006,Buzsaki2012} 
{and is now routinely used in both research and clinical diagnosis.}%and, nowadays, EEG is routinely used for both, %brain processes' research and clinical diagnosis 
\citep{invention,LopesdaSilva2010}. The oscillatory behavior observed in the different frequency bands in EEG recordings %play an important role defining the functional and effective connectivity in the 
%brain~\citep{Buzsaki2009}. Furthermore, synchronization quantifiers related with 
%these oscillations 
{ carries information about brain operation states}\citep{eeg_review_mental_2019,Babiloni2025}. %Here we will focus on alpha 
%(8-13 Hz) and beta (13-30 Hz) bands which are closely related with specific mental 
%states. Specifically, 
For instance, the power of the alpha band (8-13 Hz) increases when a person is relaxed, at rest, 
with eyes closed, and it is reduced when the person opens the eyes \citep{eeg_alpha_eo_ec_2007}. %On the 
%other hand, EEG’s alpha band analysis have been shown to be important to discern 
%illness in Alzheimer's disease~\citep{Babiloni2025}.

%~\citep{LopesdaSilva2010,LopesdaSilva2013}.
%As a result of intensive research, a whole catalog of brain states can be 
%characterized using EEG measurements. %~\citep{Rowan2003}. 
%For instance, %resting state 
%EEG recordings evolve during all the childhood or during adult 
%life~\citep{Eisermann2013,Kaminska2019,Plouin2013}. 
%EEG signals differ when 
%illness such as Alzheimer's Disease appears~\citep{Babiloni2025}. %or epilepsy~\citep{Acharya2013}). 
%The characterization and classification power of 
%EEG recordings make of it a very useful tool to study the brain processes. 
{ Among the many applications, it has been shown that} long-term intracranial human EEG recordings may enable clinically-relevant epilepsy seizure prediction \citep{epilepsy}. Besides, 
EEG-based sensors can decode neural signals and identify motor actions \citep{montesano}, which is important for the development of practical
Brain-Computer interfaces \citep{bmi}.
For instance, EEG recordings can be analyzed using machine learning algorithms and information theoretic measures to identify and characterize sleep stages \citep{sleep,rosso_sleep_pe_2021}

%Moreover, the Spatio-temporal analysis of depth-EEG signals are used in presurgical 
%techniques for drug-resistant partial epilepsy patients~\citep{Cosandier-Riml2007}.

%Electroencephalography 
%EEG has many advantages over other neurological measuring techniques (SUCH AS?). It is a low-cost non-invasive technique with high temporal resolution. 
%This characteristic makes EEG suitable to study the evolution of the voltages measured in the electrodes because, in general, the different cognitive processes involve the generation of voltage oscillations that are conventionally classified in different %production of this voltages are arranged in 
%frequency bands - delta, theta, alpha, beta, and gamma%- associated with dynamics that operate at different time scales
%~\citep{Basar2001,Kumar2012}. 

%Of course, 
While EEG is a low-cost non-invasive technique with high temporal resolution, it 
has also some disadvantages: it has poorer spatial resolution in comparison with %For instance, its spatial resolution is 
%poor in comparison to other techniques like 
magnetic resonance imaging (MRI) 
%Furthermore, 
and EEG recordings are often contaminated by electrical artifacts produced by 
%electrical 
processes such as blinking, eye movements, muscle or cardiac activity, 
etc. These %electrical 
%processes interfere strongly with the somehow weaker signals recorded from the brain in the scalp. 
artifacts can hide the signals generated by neural activity. 
Even though technicians are %can try to be very 
careful when defining the %ir experimental 
recording conditions (e.g. %making good 
carefully adjusting contacts 
between the skin and the electrodes), it is %somehow 
often unavoidable to %obtain these 
%electrical 
contaminate EEG recordings with artifacts. %So, in order to perform a sensible 
%analysis of EEG recordings, 
{Therefore, EEG raw signals are usually pre-processed to remove them.}

Detecting, rejecting and/or filtering artifacts has become, 
for many years, an active field of research producing an extensive repertoire 
of techniques that aim to remove artifacts %completely 
without distortion of relevant brain activity. %~\citep{Croft2000,Ille2002,Islam2016}.
The %type of 
approaches %created to achieve this objective is varied. Spectral 
that have been developed include spectral filters, wavelets, Principal Component Analysis (PCA) and 
Independent Component Analysis (ICA) methods among others, %are used to remove 
%artifacts
%~\citep{artifact_2007,CrespoGarcia2008} 
and new methods continue to be developed and 
combined--see~\citep{eeg_artifact_removal_review_2019} for a recent review. %In some cases, different statistical methods are 
%combined to remove, simultaneously, different artifacts~\citep{Chen2017}. 
%Synthetic (simulated) data is also used to test these methods and, sometimes, also used in the detection/rejection protocols (REF?). %Artificial Intelligence techniques are used, also, to extract statistical features or to classify brain states~\citep{Gopan2017}.

However, %nonstationary signal tools such as wavelets and time-frequency distributions are effective in characterizing transient phenomena found in event-related potentials~\citep{Aviyente2004}. So, the question arises if the rejection 
the removal of artifacts %, which are also non-stationary processes, exclude 
may also erase significant and valuable %brain 
information. %which quantifiers robust to artifacts could measure.
Traditionally, the %human 
intervention of trained technicians to identify the artifacts %epochs in the EEG recordings 
has been necessary. However, the development %linear and nonlinear analysis techniques~\citep{Stam2005} assisted by 
of computer-assisted analysis techniques has allowed the use of %to start to create 
%automatic methods
unsupervised algorithms to perform these detection/removal tasks. %, for example, during REM sleep stage~\citep{Anderer1999}. 
Nowadays, software plug-ins are routinely used to discard artifacts. %~\citep{Islam2016}. 
However, automatic methods can lead to errors (in terms of missed artifacts, or the removal of false artifacts) and thus, %spurious results. Thus, reliable artifact processing is mandatory for the correct interpretation of the EEG signals.
%Although 
%combining more than one algorithm to correct the signal in multiple processing stages is being explored as the optimal approach for artifact removal~\citep{Urigen2015}. Critical 
critical reviews of these techniques conclude that none is fully satisfactory \citep{Barban2021,Gorjan2022}. Therefore, in spite of intensive research, {EEG signal pre-processing remains challenging.}

{A possible alternative strategy {to the application of pre-processing techniques} would be to develop EEG analysis tools that do not rely on artifact removal to produce satisfactory results.}
Permutation entropy (PE)~\citep{Bandt2002}  is a popular complexity quantifier for the analysis of EEG recordings \citep{Olofsen2008,multiscale_pe_eeg_2010,QuinteroQuiroz2018,parlitz_age_pe_2021,rosso_sleep_pe_2021,Boaretto2023,klaus_review_2023,zunino_2024,Gancio2024,perinelli_2025}.
It is based on calculating the probabilities of symbols, known as {\it ordinal patterns}, that are defined by the relative values of sets of data points. Important advantages of this technique are that it is computationally efficient and robust to noise \citep{Bandt2002}. 

Here, we show that the PE is a quantity that is robust with respect to the presence of artifacts, allowing the {classification} of EEG records of healthy subjects who have their eyes closed or {their eyes} open. {This robustness is grounded in the fact the blinking artifacts act like a temporal monotonous increase of the signal, to which PE is p to\cite{amigo2010permutation,weiss2022non}.} To the best of our knowledge, PE has not yet been tested with respect to EEG artifacts.
%\comment{COMMENT: Mirad la primera frase de su sección Results de la primera página!!} or of the Alzheimer Disease. % NO LO CITO ~\cite{Puche2021}. 
%One interesting property of PE-based quantifiers is that they are still reliable when they are applied to short time series~\citep{QuinteroQuiroz2018,Gancio2024}.
%Another useful quality of PE-based quantifiers is that their temporal definition can also be 

While PE was originally proposed for time series analysis, to uncover temporal relationships between successive data values in a signal (for EEG analysis, in a single channel), it has been extended to an equivalent spatial definition~\citep{ribeiro_2012}, which allows to analyze relationships between data values recorded in different EEG channels~\citep{Boaretto2023}. In a previous work \citep{Gancio2024} several of us have used PE to discriminate the eyes-open (EO) and eyes-closed (EC) brain states from the analysis of raw EEG recordings. We found that 
the temporal and spatial PE provide complementary characterization of oscillatory brain activity %(and the brain processes involved in generating it) 
across different regions of the scalp. 
%Specifically, we found that some PE-based features provide about 75\%
%classification accuracy, comparable to the performance of features extracted with other statistical analysis techniques~\citep{Gopan2017}. \giulio{G: these other analysis were done with or without artifact removal?}
{Specifically, we found that PE-temporal features extracted from the raw data provided best accuracy, which was comparable to that obtained in~Gopan et al.\citep{Gopan2017} for the analysis of raw data, using standard statistical analysis techniques. However, we speculate that the good-performance features found in~Gopan et al.\citep{Gopan2017} mainly captured the occurrence of blinking artifacts and thus, the good classification was due to the fact that blinking artifacts only occur in the eyes-open state}. %\giulio{G: these other analysis were done with or without artifact removal?}

{To address the effect of blinking artifacts, in this work} we %continue analyzing whether PE and SPE statistical measures applied to resting EEG signals, help to 
%discriminate brain states associated with eyes open (EO) and eyes closed (EC). 
%We evaluate their reliability in terms of the number of channels used or the signal duration considered.
%We observe, in some cases, that the 
analyze the effect of pre-processing EEG recordings to manually remove the artifacts. {We find that PE quantifiers are almost unaffected by artifact removal (have a similar performance  %make the brain state discrimination 
in the raw data and in the pre-processed data), reveling that they capture genuine differences between brain states rather than the presence or absence of artifacts.}
%have a similar performance for %make the brain state discrimination 
%discriminating EO/EC states 
%in the raw data and in the pre-processed data. 
We also test the robustness of PE analysis in terms of the number of channels recorded and the duration of the recordings.

\begin{figure*}[tb] %fig 1
    \centering
    \includegraphics[width=0.8\linewidth]{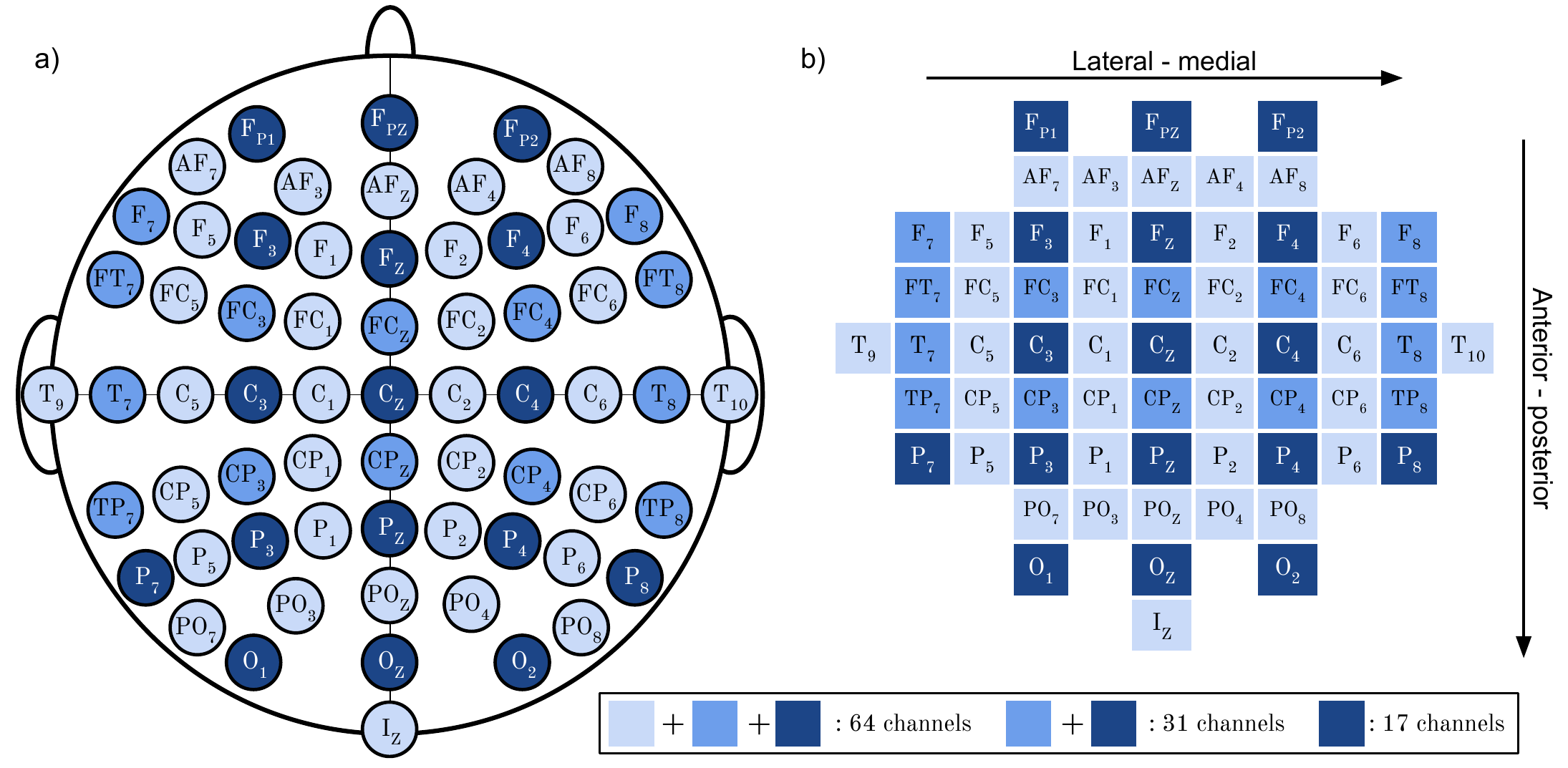}
    \includegraphics[width=0.4\linewidth]{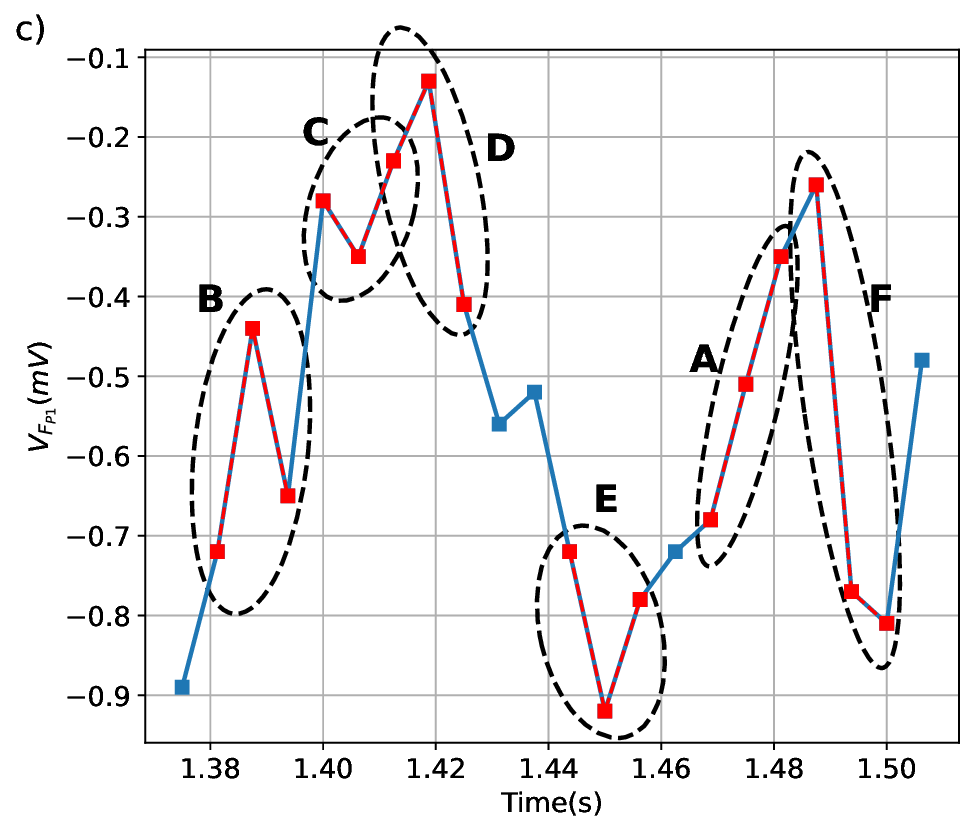}
    \includegraphics[width=0.4\linewidth]{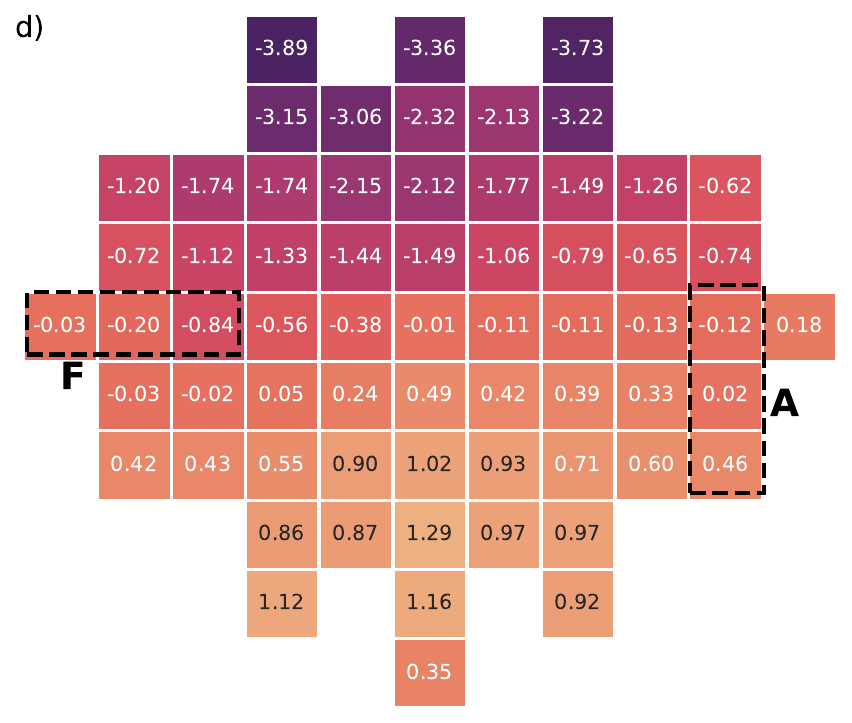}
    \caption{a) {Location of the 64 electrodes (channels) on the scalp. The inset indicates the channels that are included in the two sub-sets that we analyze (color coded): the 31 channels subset is shown in dark and dark-light blue while the 17 channels subset is shown in dark blue.} b) Grid arrangement of the electrodes. c) Examples of six temporal ordinal patterns of length $L=3$ (labeled A-F), defined over a short segment of a time series recorded in channel $F_{P1}$. d) Snapshot of the data values recorded from subject S001 at a time when a blinking occurs (values in mV and represented in color code). Here two spatial ordinal patterns are highlighted as examples: a vertical OP {of type A} and a horizontal OP { of type F}.}
    \label{fig:montage}
\end{figure*}

%\section{Materials and methods}
\section{Data} \label{sec:data}
We use a publicly available dataset of EEG recordings of 109 healty subjects \citep{schalk2004bci2000,goldberger2000physiobank}. For each subject, the dataset contains two one-minute recordings, one with eyes open (EO) and one with  eyes closed (EC). Each recording has 64 electrodes (channels), sampled at $160$ Hz, whose position is shown in Fig.~\ref{fig:montage}a. {To test the robustness of permutation entropy with respect to the number of electrodes analyzed,} we also considered two subsets of %these 
electrodes, %one 
with 31 %electrodes 
and %another 
with 17 electrodes, also {shown} in Fig.~\ref{fig:montage}a with different colors. %These subsets are %intended to emulate other montages that are commonly used; for example, the 31-channel set resembles the 

{The 31-channel subset is similar to the} one used in Liu et al.\cite{liu2024eeg}, while the 17-channel %set aims to emulate the first dataset 
{is similar to the one }
used in Quintero-Quiroz et al.\cite{QuinteroQuiroz2018}. In both cases, {we, additionally, included the $F_{PZ}$ channel to be able to define a symbol} (an horizontal ordinal pattern, as explained in section \ref{sec:ops}) with the values of three channels in the frontal region.%\textcolor{red}{ --I am curious why this channel was previously omitted.} \textcolor{blue}{No es que haya sido omitido, sino que los montajes que nos proponemos emular no lo contienen.} 

Because subjects {number 97 and 109} have slightly shorter recordings, %\citep{Boaretto2023}, we only considered 
{to homogenize the dataset, we analyzed the first $59$ seconds of all recording of all subjects; therefore, for each subject and for each state, the dataset has 64 signals with 9440 data points each.}

 \begin{figure*}[tb] %fig 2
    \centering
    \includegraphics[width=0.24\linewidth]{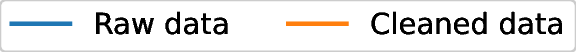} \\
    \includegraphics[width=.45\linewidth]{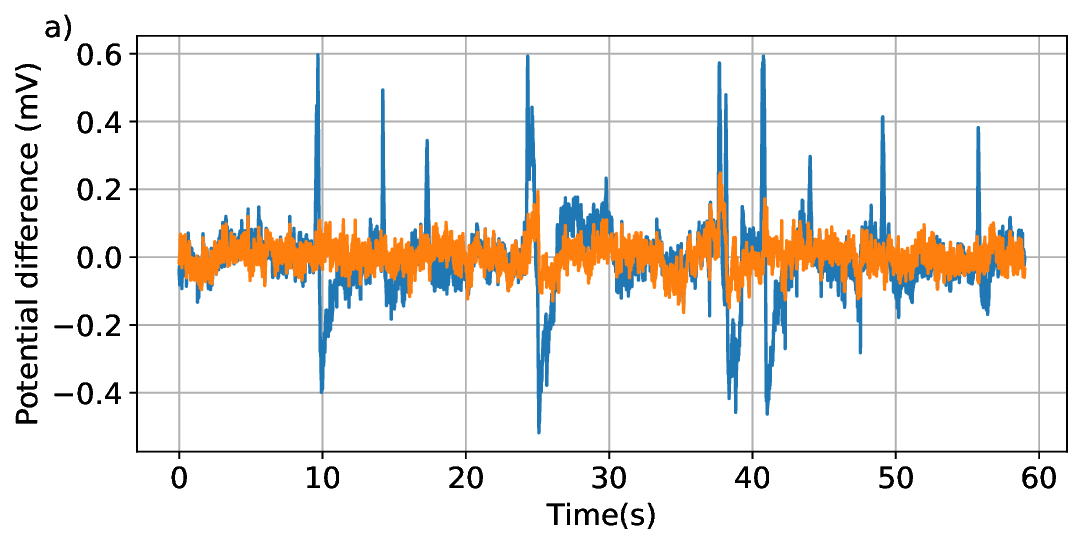}
    \includegraphics[width=.45\linewidth]{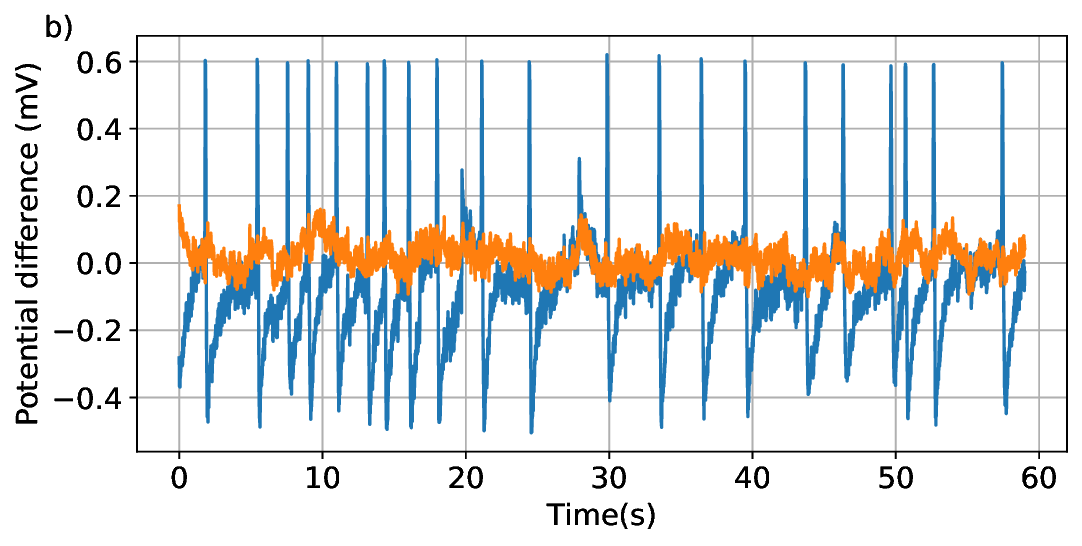}
    \includegraphics[width=.45\linewidth]{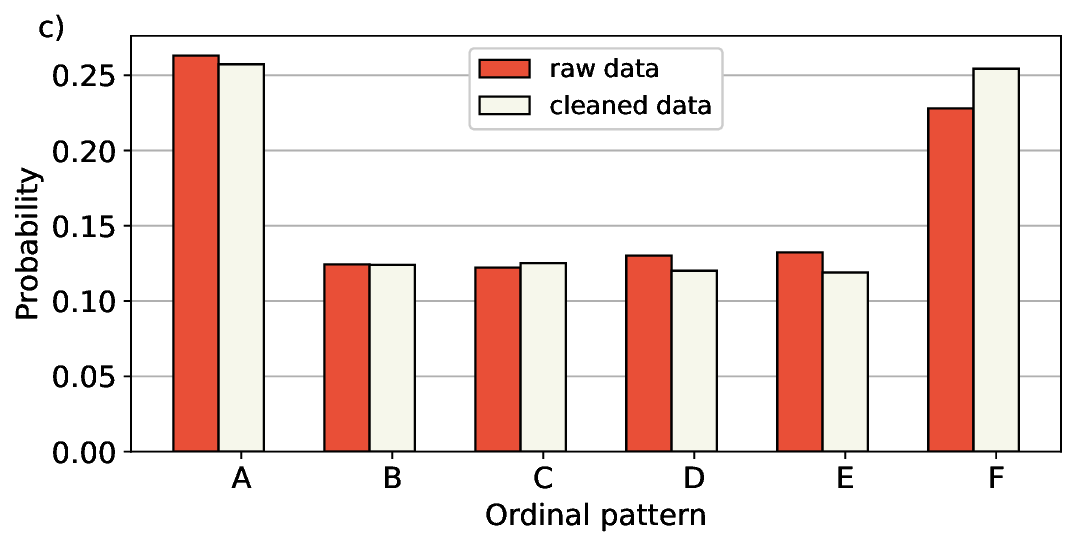}
    \includegraphics[width=.45\linewidth]{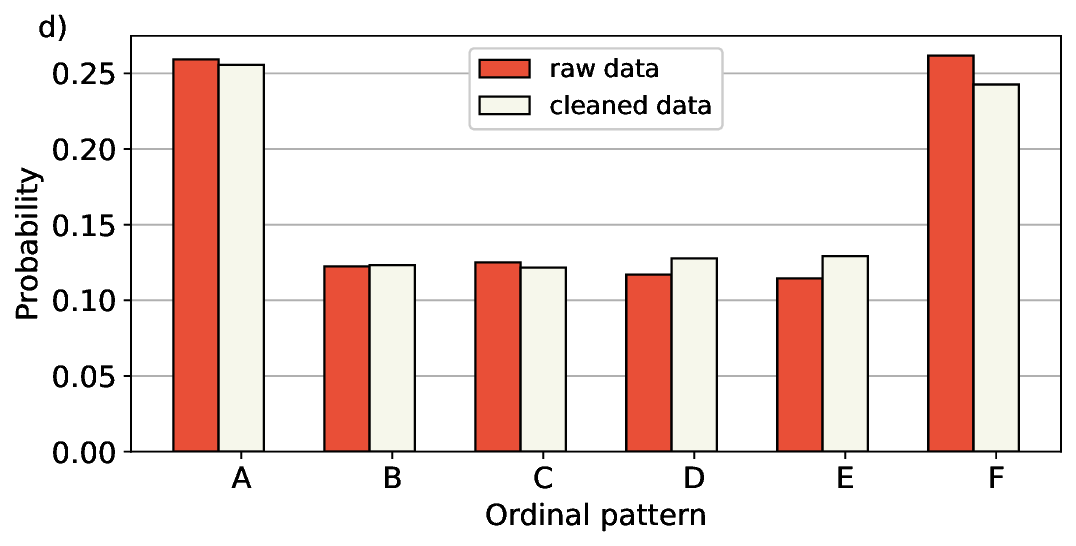}
    \caption{Examples of artifacts in time series recorded in the eyes open state, and the corresponding cleaned data. The artifacts are clearly noticed as abrupt and out-of-scale jumps in the signal. Panel a) displays the time series recorded in channel $F_{P1}$ of subject S001 that contains 9 artifacts while panel b) displays the recording in channel $F_{P2}$ of subject S003 that contains 21 artifacts. {Panels c) and d) show the probability distributions of the OPs obtained from the time series shown in panels a) and b), respectively. From these distributions, the $PE$ obtained from raw data is $0.970$ (panel a) and $0.958$(panel b), and from cleaned data is $0.962$ (panel a) and $0.968$ (panel b), representing a relative change of $0.825\%$ (panel a) and $-1.04\%$ (panel b).}
    %Permutation entropy analysis of the signals obtained from electrode $F_{P1}$ during a blinking. In panel a), a segment of the time series of $2,5s$ of duration during a blinking artifact is displayed from the raw and cleaned data. Panel b), shows the resulting probability distribution of symbols obtained from these segments.\textcolor{red}{no se si vale la pena mostrar esta fig. Sugerencia: mostrar un par de ejemplos de raw y filtered signals y poner en el caption las entropias, sin mostrar las posibilidades. esta fig. tendria que ir en la Sec. 2.2 Removal of blinking artifacts.}
    }
    \label{fig:artifacts}
\end{figure*}

\subsection{Removal of blinking artifacts}
To test the robustness of permutation entropy to the presence of blinking artifacts, we analyzed the raw signals, and also, the post processed signals, after artifacts were removed. To remove them, we performed an independent component decomposition, or ICA analysis \citep{mantini2007complete,Barban2021}, implemented using the function provided by the MNE \emph{Python} package \citep{gramfort2013meg}, removing the independent component containing the blinking artifacts, and reconstructing the data again. %For better results, we performed this 
The procedure was performed manually for each subject \citep{hamal2013artifact}, identifying the blinkings as the independent component that presents oscillations at a characteristic %an appropriate 
periodicity of about $3.5$ s \citep{bentivoglio1997analysis} (about $17$ blinking artifacts per minute).

Two examples are shown in Fig.~\ref{fig:artifacts}, before (raw signals) and after artifact removal {for two subjects displaying different artifacts frequencies}. As it can be seen, blinking artifacts manifest as abrupt changes in the signal, with an amplitude much larger than the background activity \citep{uriguen2015eeg}. {Their spatial distribution is not homogeneous, as they} occur mainly in recordings of EEG channels that are located in the frontal region of the scalp \citep{Croft2000}

%(\textcolor{red}{which ones? rate in seconds?? or characteristic periodicity?)}. \textcolor{red}{Assuming this characteristic rate, $17$ blinking artifacts can be expected in one-minute recordings. 
%SI PERO CUANTOS SACARON? SON SIMULTANEOUS? OCURREN AL MISMO TIEMPO EN TODOS LOS CANALES? TODOS LOS CANALES TIENEN, EN PROMEDIO, EL MISMO NUMERO DE ARTEFACTOS?} \textcolor{blue}{Los artefactos son un efecto "global" en el sentido que se aislan en un componente independiente que en principio contiene contribuciones de todos los canales, obviamente esto no es así y la contribución de los canales posteriores puede ser cero, pero el alcance del artefacto depende sujeto a sujeto. Lo que si, los artefactos se eliminan simultáneamente para todo los canales. Cuando eliminamos los artefactos no contamos cuantos sacamos, ya que ni siquiera requería que observaramos toda la señal, mirabamos los IC para encontrar el que tenia contribuciones princpalmente frontopolares, y unos 20s de la señal para confirmar que presentaba oscilaciones con un período similar al que buscabamos. Si queremos saber cuantos artefactos se removieron exactamente, tendría que contarlos.} Because blinking artifacts only occur when subjects have their eyes open, the removal procedure {was only performed on EO signals. } 

\section{Permutation entropy}
The permutation entropy ($PE$)\citep{Bandt2002} is a well-known analysis tool for biomedical data \citep{review_2022}. In order to compute it for {a sequence of $N$ ordered }%a time series of $N$ 
data points, 
%\begin{equation}
$x = \{x_1,~x_2,~\dots,~x_N\}$,
%\end{equation}
we first encode it into a sequence of symbols, known as {\it ordinal patterns} (OPs) $s=\{s_1,s_2,\dots,s_n\}$, with $n=N-(L-1)$, where $L$ is {the} number of data points used to define the symbol, known as the length of the ordinal pattern. The OPs are defined in the following way: {using a sliding window,} for each $i=1,\dots,n$, the vector $[x_i,~x_{i+1},~x_{i+2},~\dots,~x_{i+(L-1)}]$ is reordered such that $x_{\pi(i)}\leq x_{\pi(i+1)}\leq\dots\leq x_{\pi(i+(L-1))}$, where $\pi(\cdot)$ is the permutation index that transforms the original vector into the ordered one. The corresponding ordinal pattern is $s_i=[\pi(i),~\pi(i+1),~\dots,~\pi(i+(L-1))]$, which corresponds to one of $L!$ possibilities. {To implement this encoding algorithm, we started from }the Python program included in \cite{Parlitz2012}, and we updated it to work in \emph{Python3}. {To break the ties formed by equal values, we assign the pattern following their order of occurrence, which is a common practice in other popular implementations\cite{pessa2021ordpy}. This has the drawback of introducing a bias in the patterns' probabilities, which, however, does not affect our findings. In fact, we analyze the differences of the entropy values, and the bias would affect in the same way the ordinal probabilities in the two states, leaving the differences unaltered\cite{cuesta2018patterns}.}

The probabilities of occurrence of the $L!$ possible ordinal patterns, $\{p_1, p_2, \dots, p_{L!}\}$ are estimated from the frequency each pattern occurs in the sequence $s= \{s_1,s_2,\dots,s_n\}$.  With the patterns' probabilities, the normalized permutation entropy is finally calculated as:
\begin{equation}
    PE=-\frac{1}{\log\left(L!\right)}\sum_{m=1}^{L!}p_m\log\left(p_m\right), \label{eq:pe}
\end{equation} 
which %verifies that $PE\in[0,1]$. 
{is in the range $[0,1]$, being equal to 0 when all the probabilities are 0 except one that is equal to 1 ($p_j=1$ and $p_k=0$ $\forall$ $k\ne j$), and being equal to 1 when all the probabilities are equal ($p_j=1/L!$ $\forall$ $j$)}.

\subsection{Temporal and spatial ordinal patterns} \label{sec:ops}

Although originally conceived to capture the temporal variations of a time series %signal 
\citep{Bandt2002}, OPs can also be {extended to} spatial data \citep{ribeiro_2012}. %in space
%, which allows us to obtain the spatial permutation entropy, $SPE$, in the same manner as in Eq.~\ref{eq:pe}.
{We refer to OPs computed from spatial data as \emph{spatial ordinal pattern}, as opposed to \emph{temporal ordinal patterns}, that are OPs computed from encoded timeseries}. To define the spatial OPs, we use $L$ channels with %consider two different orientations of the symbols: those with a preferred 
either lateral-medial orientation (referred to as {\it horizontal OPs}) or 
%others with a preferred 
anterior-posterior orientation (referred to as {\it vertical OPs}). To do this, we  %transform the positioning of the electrodes into 
{map the electrodes locations to the regular grid} shown in Fig.~\ref{fig:montage}b, and then define the ordinal patterns using values along a row (for horizontal OPs) or along a column (for vertical OPs) {of the grid}. To fix the ideas, examples of the six temporal OPs of $L=3$, labeled A-F, are shown in Fig.~\ref{fig:montage}c, and examples of an horizontal and a vertical OP are shown in Fig.~\ref{fig:montage}d.

{Finally, we refer to $PE$ as the normalized permutation entropy calculated by Eq. (\ref{eq:pe}) from the probabilities of temporal OPs, whereas we refer to spatial permutation entropy ($SPE$) as the normalized permutation entropy, also calculated with Eq. (1),}
\begin{equation}
\begin{aligned}
    SPE_H=-\frac{1}{\log\left(L!\right)}\sum_{m=1}^{L!}p_m^H\log\left(p_m^H\right),\\ 
    SPE_V=-\frac{1}{\log\left(L!\right)}\sum_{m=1}^{L!}p_m^V\log\left(p_m^V\right), \label{eq:spe}
    \end{aligned}
\end{equation} 
but using probabilities referred to spatial OPs:{ $p_m^H$ (for horizontal OP) and $p_m^V$ (vertical OP).} 

For defining temporal patterns in a given channel, $N=9440$ data points are available, while for defining spatial patterns at a given time, about $N=64,\;31 \text{ or } 17$ data points are available, depending on the orientation of the pattern and the set of channels considered (see Table~\ref{tab:number_of_patterns}). {For example, for the subset of 17 channels (marked with dark blue in Fig.~\ref{fig:montage}b) and $L=3$, we can define only one horizontal pattern in the first three rows and in the last row, while in the fourth row, 3 horizontal patterns are obtained, adding to $1 +1+1+3+1=7$ horizontal patterns. Regarding the columns, 3 vertical patterns can be defined in each of the three central columns, while no vertical pattern can be defined in the left and right columns (because they contain only one channel). Therefore, $3 \times 3 =9$ vertical patterns can be defined in this subset.} %\textcolor{blue}{(En realidad para la parte espacial se analizan filas y columnas que tienen un $N$ menor)} }

{Because we have at most 64 channels, we restrict the analysis to spatial OPs of length $L=3$, since in this case we only have to estimate the probabilities of 6 patterns. However, as shown in Table~\ref{tab:number_of_patterns},  the number of spatial OPs that can be defined is low  or very low for a precise estimation of the 6 probabilities, because in the best scenario we only have $45$ patterns. %Table~\ref{tab:number_of_patterns} displays the number of spatial OPs that can be constructed, depending on the subset of channels and the OP orientation. 
In this sense, we consider that the set of 64 channels provides a slightly under-sampled estimation of the 6 probabilities, while the subsets of 31 and 17 channels provide a moderately and severely under-sampled estimation, respectively.}

To perform a fair comparison of the performance of the temporal and spatial OPs, we also define temporal OPs of length $L=3$. Nevertheless, we found very similar results with temporal patterns of length $L=4$.

\begin{table}[tb]
    \centering
    \begin{tabular}{|c|c|c|}
    \hline
        { Number of channels} & { Number of spatial ordinal patterns} \\
        \hline
        64 &  45 Horizontal; 44 Vertical\\
        31 &  17 Horizontal; 21 Vertical \\
        17 &   7 Horizontal; 9 Vertical\\\hline
    \end{tabular}
    \caption{Number of spatial OPs that can be defined {with $L=3$}, depending on the subset of channels and on the orientation of the patterns.}
    \label{tab:number_of_patterns}
\end{table}

\subsection{{Averaging procedure}}

{For each subject $j$ ($j=1 \dots 109$), we calculated the average of $PE$ over the channels, $\langle PE\rangle^j$, and the average of $SPE$ over time, referred to as $\langle SPE\rangle^{j}_H$ and $\langle SPE\rangle^{j}_V$, when horizontal or vertical OPs were used, respectively. In this way, for each subject $j$, we obtain three average entropies in the eyes closed state, $\langle PE\rangle^{j, EC}$, $\langle SPE\rangle^{j, EC}_H$ and $\langle SPE\rangle^{j, EC}_V$, and three in the eyes open state, $\langle PE\rangle^{j, EO}$, $\langle SPE\rangle^{j, EO}_H$ and $\langle SPE\rangle^{j, EO}_V$. }%\textcolor{red}{comment cris: aqui hecho a faltar un comentario sobre "pooling", es decir, en lugar de promediar las entropias, calcular una unica entropia promediando las probabilidades.}

\subsection{Statistical significance analysis}\label{sec:pvalues}
%In order to try to distinguish EO and EC states from differences in the distributions of values of permutation entropy, 
%{The $\langle PE\rangle$ and $\langle SPE\rangle$ values obtained for the 109 subjects define three distributions for each state (the distributions of $\langle PE\rangle^{j,EO}$, $\langle SPE\rangle^{j,EO}_H$ and $\langle SPE\rangle^{j,EO}_V$ and $\langle PE\rangle^{j,EC}$, $\langle SPE\rangle^{j,EC}_H$ and $\langle SPE\rangle^{j,EC}_V$ with $j=1, \dots, 109$).
%To identify statistically significant differences, we performed a paired t-test (implemented using the Python SciPy library SciPy \citep{2020SciPy-NMeth}) to evaluate the differences between distributions in different states}. In this test, the null hypothesis (n.h.) is that the distributions have equal means. The n.h. was rejected when the p-value was~$<10^{-3}$.
{To assess whether the values of the permutation entropies for a given subject are significantly different in the EO and EC states, we performed a paired t-test (implemented using the Python SciPy library SciPy \citep{2020SciPy-NMeth}). In this test, the null hypothesis is that there are no differences between the two states, therefore the distributions of $\langle PE\rangle^{j,EC} - \langle PE\rangle^{j,EO}$ and $\langle SPE\rangle^{j,EC} - \langle SPE\rangle^{j,EO}$ have 0 mean. The null hypothesis was rejected for p-values less than $10^{-3}$.}

\section{Results}

\begin{figure}[tb] %fig. 3
    \centering
    \includegraphics[width=.95\linewidth]{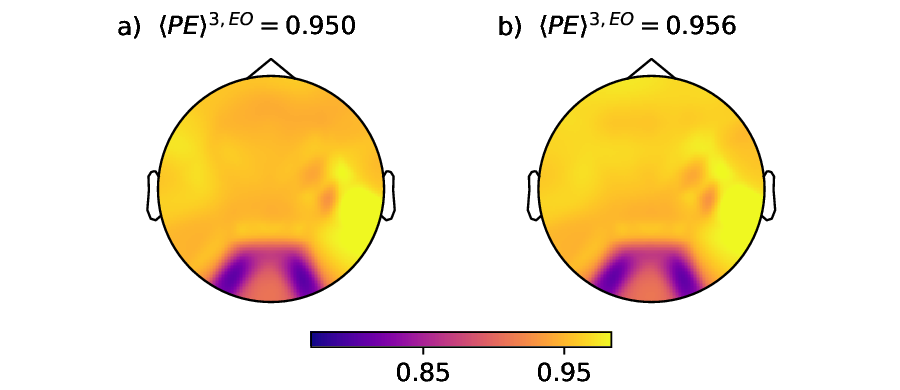}
    \includegraphics[width=.95\linewidth]{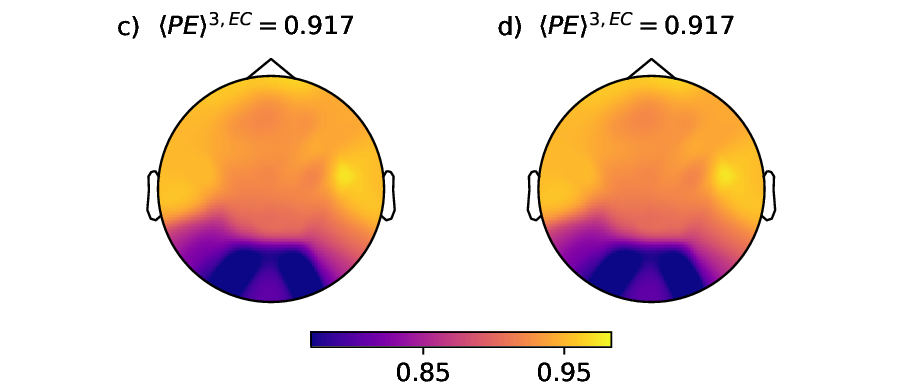}
    \includegraphics[width=.95\linewidth]{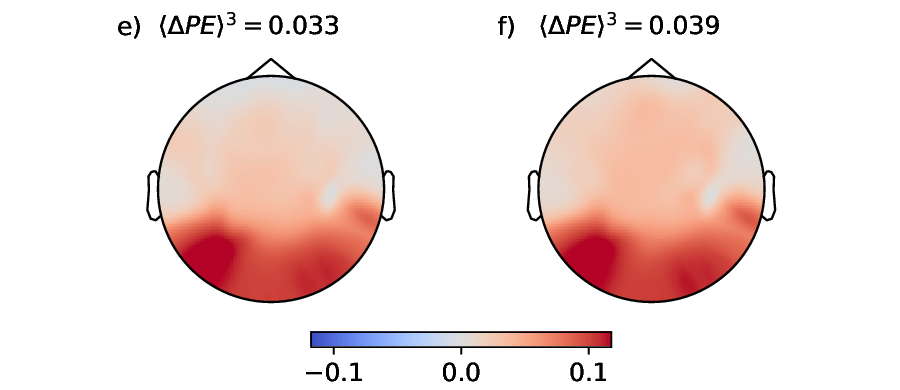}
    \caption{Topographic maps of permutation entropy ($PE$) calculated for the 64 channels of a single subject (S003). Left column corresponds to the analysis of the raw data, while right column, to the analysis of the filtered data, where 21 blinking artifacts were removed. Panels a) and b) show $PE$ values in the EO state ($PE^{3,EO}_k$ with $k=1\dots 64$), c) and d) $PE$ in the EC state, $PE^{3,EC}_k$ (these two panels are identical because artifacts occur only in EO signals) and e) and f) display the difference, $PE^{3,EO}_k-PE^{3,EC}_k$. The average values of $PE^{3,EO}$, $PE^{3,EC}$ and $PE^{3,EO}-PE^{3,EC}$ over the 64 channels are indicated, and we see that the values are almost the same in a) and b) and in e) and f).}
    \label{fig:topomaps_PE}
\end{figure}

\subsection{{Effect of artifact removal}}

We begin by presenting the analysis of a single subject, S003, that in the EO state presents a large number of artifacts, as shown in Fig.~\ref{fig:artifacts}(b). {Figure~\ref{fig:topomaps_PE} displays the temporal permutation entropy for the raw data (left column) and for the filtered data (right column). Here the color code indicates %, calculated from $L=3$ temporal OPs, , 
the temporal permutation entropy for eyes opened (top panels), for eyes closed (center panels) and the difference between the two (bottom panels) for all the 64 channels.}
Because artifacts are removed only in the EO state, Figs.~\ref{fig:topomaps_PE}c and \ref{fig:topomaps_PE}d are the same. 
%, displayed on a topographic map of the scalp. 
%In this figure, the top row displays the entropy values of the raw signals (no removal of the blinking artifacts), while the bottom row, of the pre-processed signals (after artifact removal). %Panels a--c of 
The entropy values %averaged over the 64 channels are also indicated and }
%Fig.~\ref{fig:topomaps_PE} %resemble the analysis 
are consistent with those 
reported in \cite{QuinteroQuiroz2018}, where %we observe that the 
EC states %report
were found to have 
lower entropy in all the channels, especially in the posterior region over the visual cortex, where the difference between EC and EO is largest. This is observed in the raw data and also in the filtered data (after artifact removal).

%Panels d--f of Fig.~\ref{fig:topomaps_PE} present our new analysis, where the blinking artifacts have been removed. Because these artifacts occur only when the subjects have their eyes opened, these do not occur during EC states, thus 
Remarkably, {almost no difference is observed between Figs.~\ref{fig:topomaps_PE}a and \ref{fig:topomaps_PE}b or between Figs.~\ref{fig:topomaps_PE}e and \ref{fig:topomaps_PE}f}, which is a first indication that permutation entropy analysis is robust to the presence of blinking artifacts.%Regarding panels d and f, where the removal of the artifacts could imply differences in the analysis, we do not observe any significant change in the results, and no changes in the frontal region of the scalp, where the effects of the blinking artifacts are expected. This indicates that permutation entropy is quite robust to this kind of artifacts. 

The robustness of the permutation entropy analysis %to the blinking artifacts is based on 
is due to the core feature of this technique: the ordinal patterns are not defined in terms of % does not depend on 
the absolute values of %neighboring 
the data points in the signal, but %only 
on their relative values. %variation of them. These artifacts are captured mainly by the frontal electrodes of the EEG, which substantially alter their values during blinking \citep{QuinteroQuiroz2018}. %The unaffectedness of PE implies that these drops only act as temporal addition of a constant to the signal, keeping its local structure of relative amplitudes. 
The fact that $PE$ is almost unaffected by artifact removal suggests that the procedure used to remove artifacts {maintains the statistical distribution of the ordinal patterns. Although the sequence of OPs obtained from the raw and from cleaned time series in the pre-frontal region is not the same (only $40\%$ of the time the OP that occurs in the raw signal and in the filtered signal is the same), the differences compensate and the resulting OP probability distributions, shown in Fig.~\ref{fig:artifacts}c,d,  are very similar.}

\begin{figure}[tb] %fig 4
    \centering
    \includegraphics[width=0.75\linewidth]{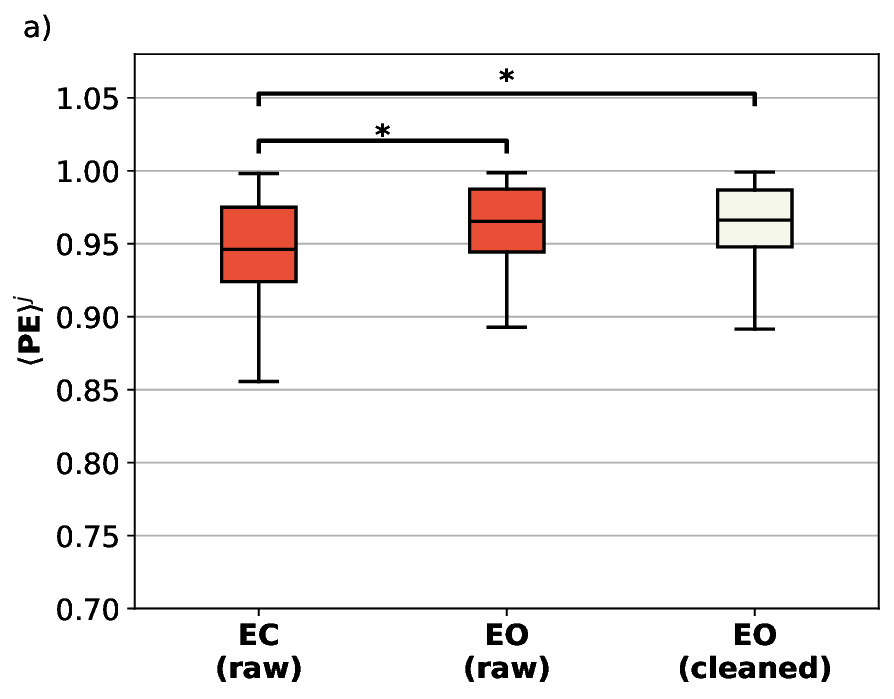}
    \includegraphics[width=0.75\linewidth]{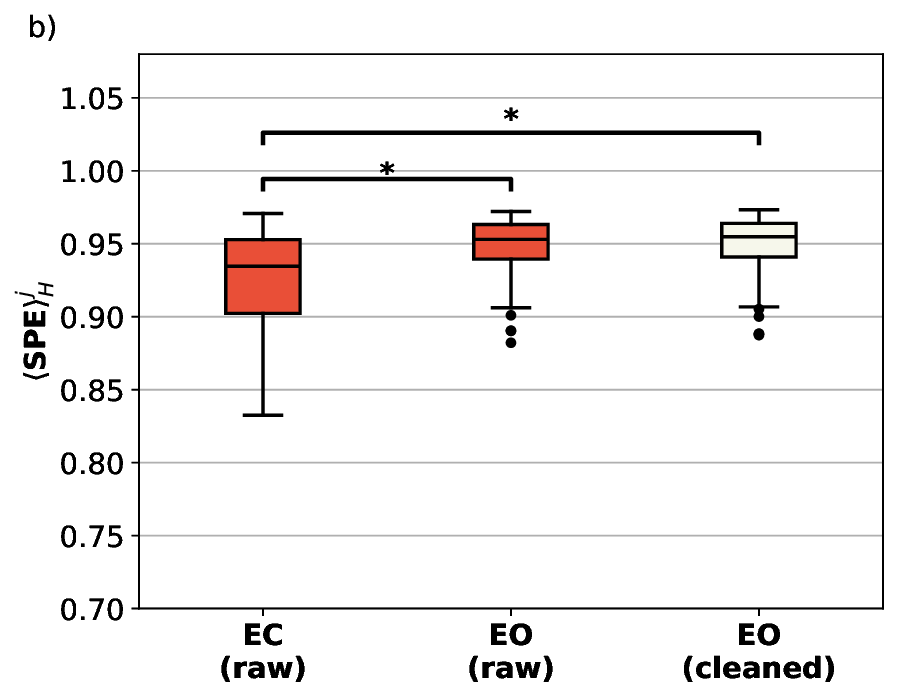}
    \includegraphics[width=0.75\linewidth]{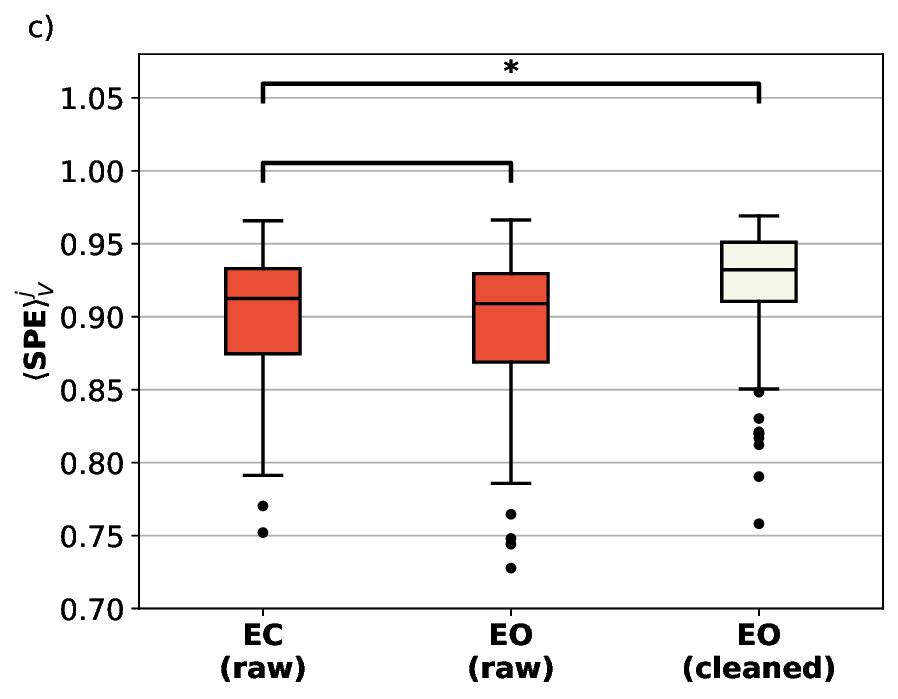}
    \caption{{Distributions of average entropy values obtained from all subjects, in the EO and EC states}, from raw data (red) and from artifacts removed data (white). (a) $\langle PE\rangle^j$ values; (b) $\langle SPE\rangle^j_H$ values, and (c) $\langle SPE\rangle^j_V$ values. %Panel a) shows de distribution of PE values, panel b) the one corresponding to SPE with horizontal symbols, and panel c) to SPE with vertical symbols. 
    Symbol * indicates a p-value~$<10^{-3}$.}
%    \giulio{G: In this figure you heve to increase the upper limit of the vertical axis or the "*" touch the upper limit. (you can fix c bu lowering the horizontal [ but panel a) is already at the limit.}
    \label{fig:PE_boxplot}
\end{figure}

We calculated $PE$ and $SPE$ values in the EC and EO states (with and without blinking artifacts) for the recordings of all subjects.
Figure~\ref{fig:PE_boxplot} displays the distributions of $\langle PE\rangle^{j,EO}$, $\langle SPE\rangle^{j,EO}_H$, $\langle SPE\rangle^{j,EO}_V$, $\langle PE\rangle^{j,EC}$, $\langle SPE\rangle^{j,EC}_H$, and $\langle SPE\rangle^{j,EC}_V$, with $j=1, \dots, 109$. %For the EO state, entropy obtained from recordings with and without artifacts are presented.
For $\langle PE\rangle^{j,EO}$ (Fig.~\ref{fig:PE_boxplot}a) no differences with or without artifacts can be observed. In this case, the difference between EO and EC states is significant (p-value $<10^{-3}$), regardless artifact removal.
%When the permutation entropy is calculated from the probabilities of horizontal OPs,
Also in Fig.~\ref{fig:PE_boxplot}b, %the symbols used have a horizontal orientation, and 
we see that the removal of artifacts has little impact on the distribution of %$\langle SPE\rangle$ 
$\langle SPE\rangle^{j,EO}_H$
values, and the difference between EO and EC states is significant %also maintains the significance level 
(p-value $<10^{-3}$). Again, we attribute this robustness to the same reason as before: artifacts do not change the relative horizontal spatial structure of the data values. %seem to affect the data points that made up the symbol equally (as a constant added to the signal). 

%Now we present our 
%Next, we present the results %regarding the spatial approach of permutation entropy, 
%obtained when using the spatial approach to define the ordinal patterns.
%Spatial Permutation Entropy (SPE). 
%Fig.~\ref{fig:PE_boxplot}b-c shows the distribution of the average SPE of each subject, $\langle SPE\rangle$, for the different brain states, but also using different symbol constructions. 

However, %when the permutation entropy is computed from the probabilities of vertical OPs, 
Fig.~\ref{fig:PE_boxplot}c shows %the results when the symbols are constructed with vertical orientation. In this case, the 
that artifacts do affect the $SPE_V$ values: %that once removed, we recover the usual behavior of 
{after artifacts are removed, the $SPE_V$ of EO states increases with respect to $SPE_V$ calculated from the raw data, and it becomes larger that $SPE_V$ of EC states, i.e., in the filtered data $SPE_V(EO)> SPE_V(EC)$.}
%after artifacts are removed, EO states have higher entropy than EC states (as occurs for $PE$ and for $SPE_H$). 
Because some vertical OP are defined by data values recorded in electrodes in the region affected by the artifacts (the first row of frontal electrodes: $F_{P1}$, $F_{PZ}$, and $F_{P2}$), and also, in electrodes outside this region, and due to the slow recovery of the signal after a blinking (as shown in Fig.~\ref{fig:artifacts}), data values recorded in electrodes affected by artifacts tend to be lower than the data values recorded in other electrodes (as shown in Fig.~\ref{fig:montage}d). Therefore, in the raw data, the vertical OP that encodes a positive trend (three increasing values such as 1-2-3) is over-expressed and the OP that encodes a negative trend (three decreasing values such as 3-2-1) is under-expressed with respect to the vertical OPs that encode other relative ordering (e.g., 2-1-3, 3-2-1, etc.). 

\begin{figure}[tb] %fig5
    \centering
    \includegraphics[width=0.75\linewidth]{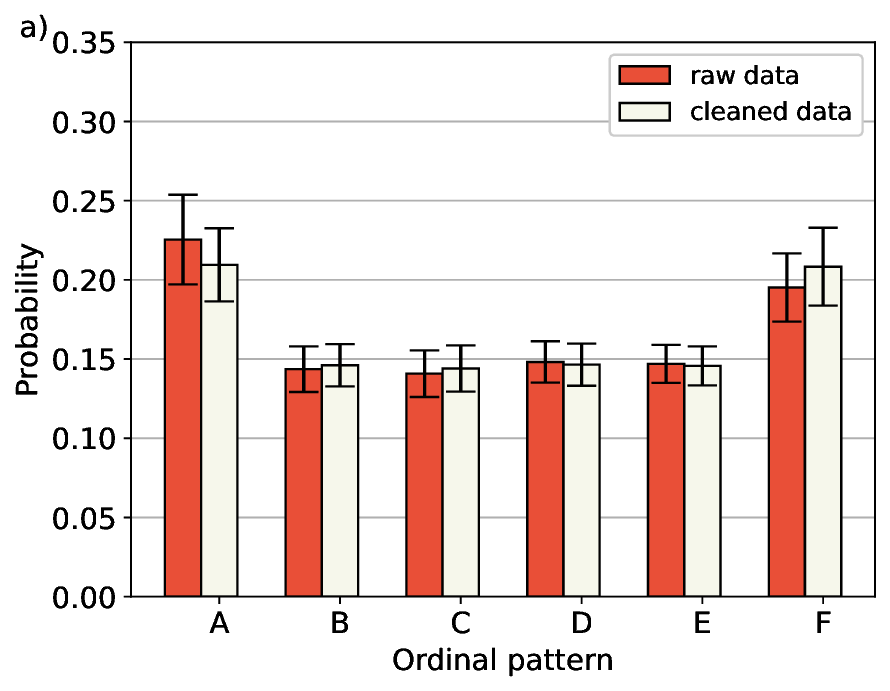}
   \includegraphics[width=0.75\linewidth]{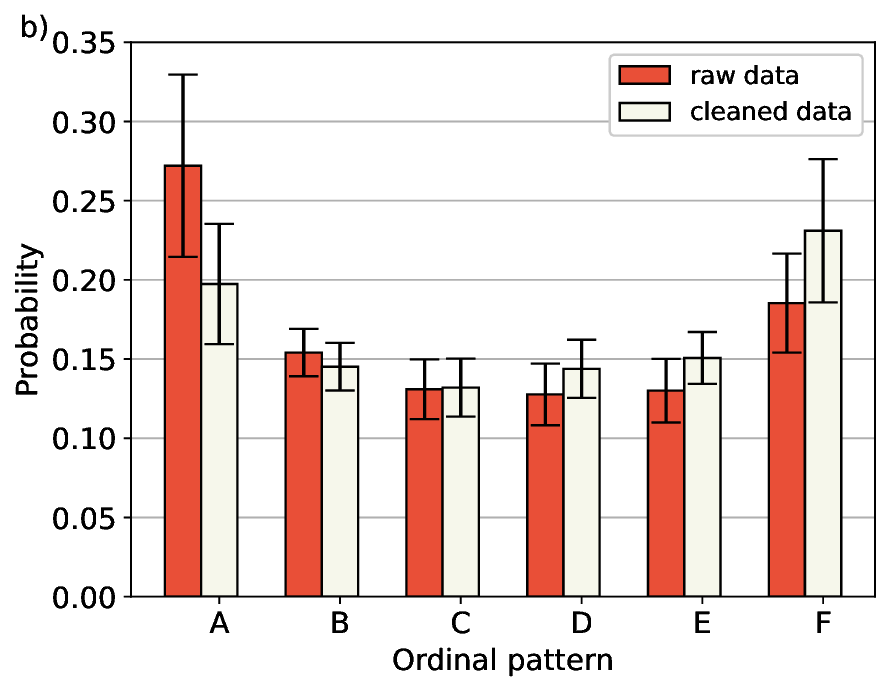}
    \caption{Probabilities of spatial ordinal patterns for the EO states, from raw data and from artifacts removed data, averaged over all subjects (errors bars indicate $\pm1$ standard deviation). Panel a (panel b) corresponds to horizontal (vertical) OPs. Artifact removal produces small variations of the probabilities of the horizontal OPs, while it produces larger variations of the probabilities of the vertical OPs, decreasing the probability of $A$ (the patterns that encodes a positive trend) and increasing the probability of $F$ (the pattern that encodes a negative trend).}
    \label{fig:SPE_probs}
\end{figure}

To confirm this interpretation, Fig.~\ref{fig:SPE_probs} displays the probabilities of the $6$ horizontal OPs (panel a) and vertical OPs (panel b), obtained from raw data (red) and cleaned data (white), averaged over all subjects. Indeed, for the horizontal OPs, the probabilities are very similar in raw and cleaned data. In contrast, for vertical OPs, in the raw data A pattern (1-2-3) is over-expressed while F pattern (3-2-1) under-expressed, while the opposite occurs in the cleaned data.%symbols from this frontal region predominantly of the structure type of continuously increasing magnitude (symbol A, see Fig.~\ref{fig:montage}c). %PARA NO CONFUNDIR --AQUI HABLAMOS DE OP VERTICAL Y EN LA FIGURA MOSTRAMOS OP TEMPORAL --SUGIERO OMITIR ESTA FRASE.

Therefore, for horizontal OPs there is no significant difference in the probabilities before or after removing the artifacts, but for vertical OPs, the probability of pattern A is higher when analyzing the raw signals with artifacts. This effect lowers $SPE^{k,EO}_V$ and make its distribution very similar to the distribution of $SPE^{k,EC}_V$ (Fig.~\ref{fig:PE_boxplot}c).
%This is confirmed in Fig.~\ref{fig:SPE_probs}, where we show the probability distribution of the 6 possible symbols (for $L=3$), with and without artifacts. For the horizontal orientation, Fig.~\ref{fig:SPE_probs}a, there is no significant difference in the probability distribution before or after removing the artifacts. But for the vertical orientation, Fig.~\ref{fig:SPE_probs}b, the probability of the symbol A is significantly greater when analyzing the signals with artifacts. This lowers the values of SPE and makes the distribution of EO states very similar to the EC states (red boxplot). 
{We believe this is the reason why the vertical orientation was found to perform } %This is the reason that this symbol orientation performs 
poorly in Gancio et al.\cite{Gancio2024}, but it performed well when analyzing only the alpha band, as this band does not contain blinking artifacts.

\begin{figure}[tb] %fig 6
    \centering
    \includegraphics[width=0.45\linewidth]{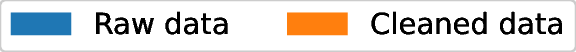}
    \includegraphics[width=0.75\linewidth]{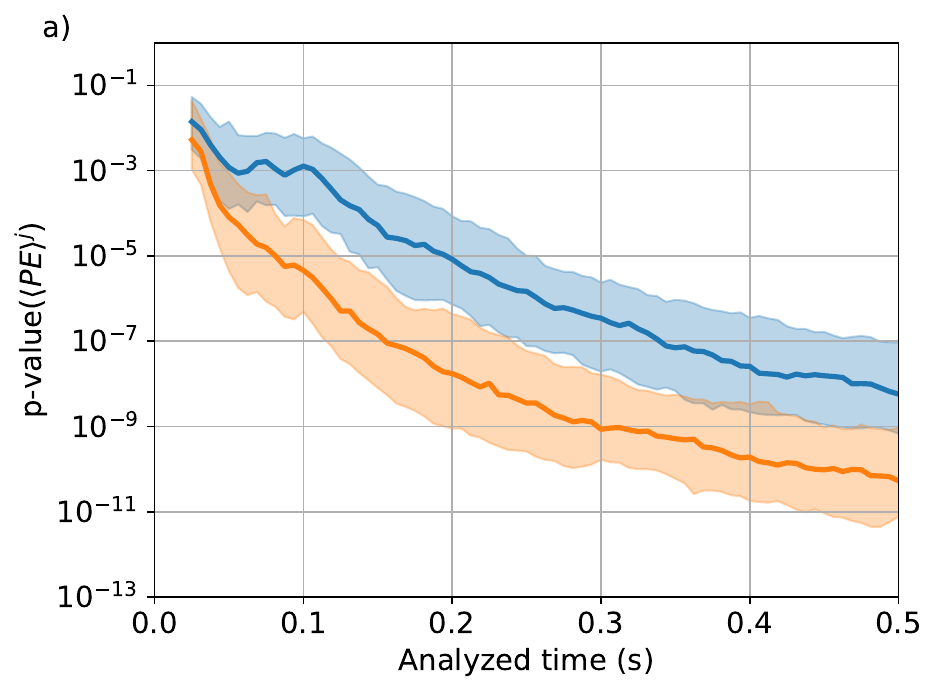}
    \includegraphics[width=0.75\linewidth]{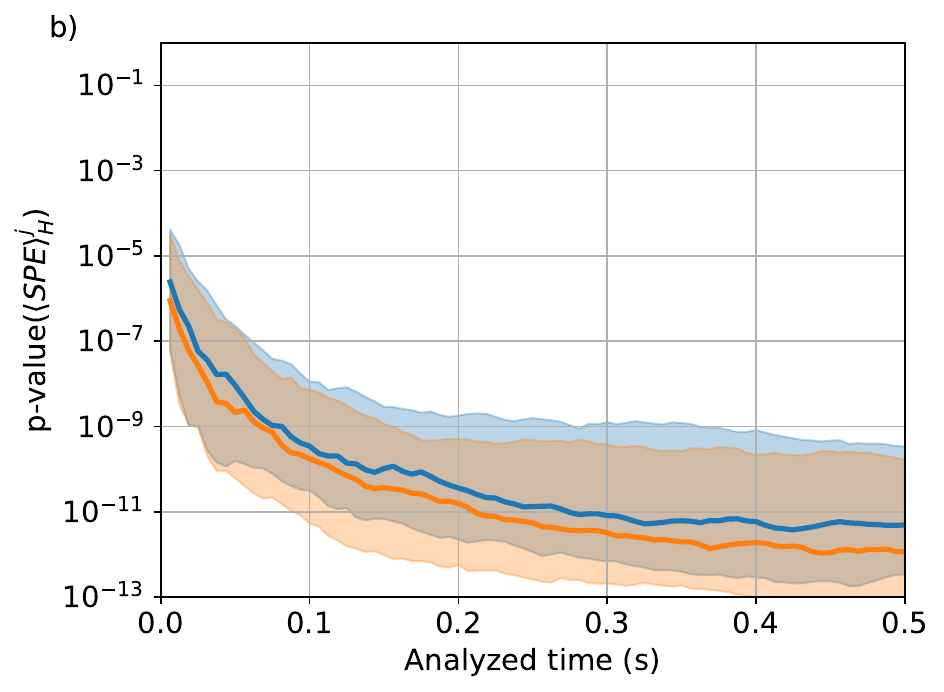}
    \includegraphics[width=0.75\linewidth]{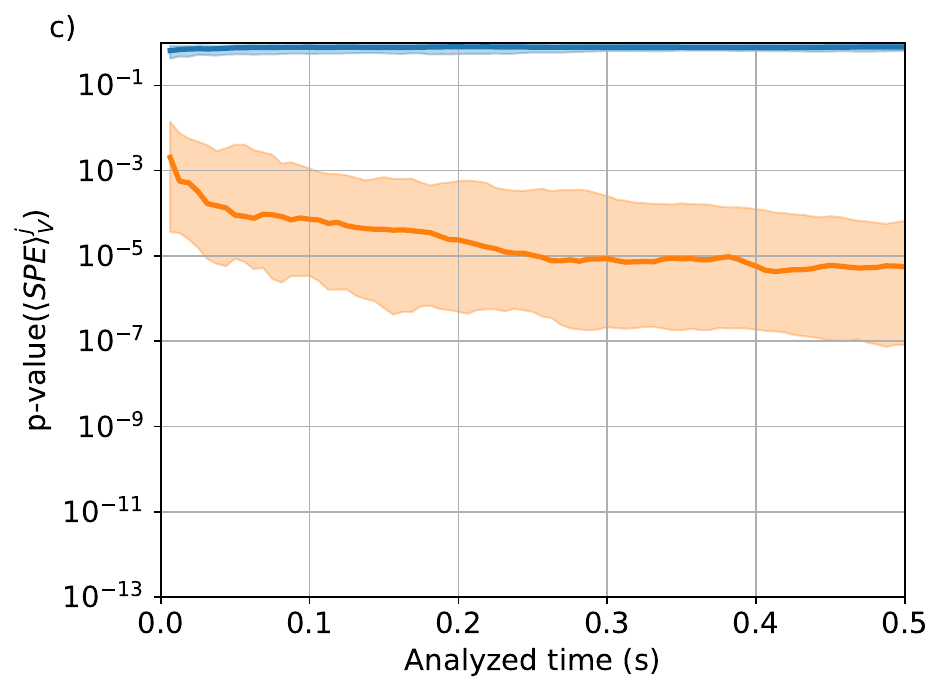}
    \caption{{P-values obtained from the comparison of the entropy distributions (a: $\langle PE\rangle^{j}$, b: $\langle SPE\rangle^{j}_H$, c: $\langle SPE\rangle^{j}_V$) in the EO and EC states, calculated in 118 non-overlapping segments of $0.5$~s each, as a function of the time interval analyzed in each segment. The continuous line shows the median of the 118 p-values, while shaded areas indicate the spread between first and third quartiles. 
    %In panel c, the data is not accumulated in time, but analyzed using a sliding window of $0,5s$ width, and the p-values are shown as a function of the starting time of such window, measured from the beginning of the experiment ($t=0$). And in panel d, we shuffled the SPE values in time, and plot the p-values as a function of the number of snapshots used to calculate the average SPE per subject. 
    %Continuous (dashed) lines indicate p-values obtained from the analysis of recordings with (without) blinking artifacts. 
    }}
    \label{fig:pvalues_as_t}
\end{figure}

\begin{figure}[tb] %fig. 7
    \centering
    \includegraphics[width=0.75\linewidth]{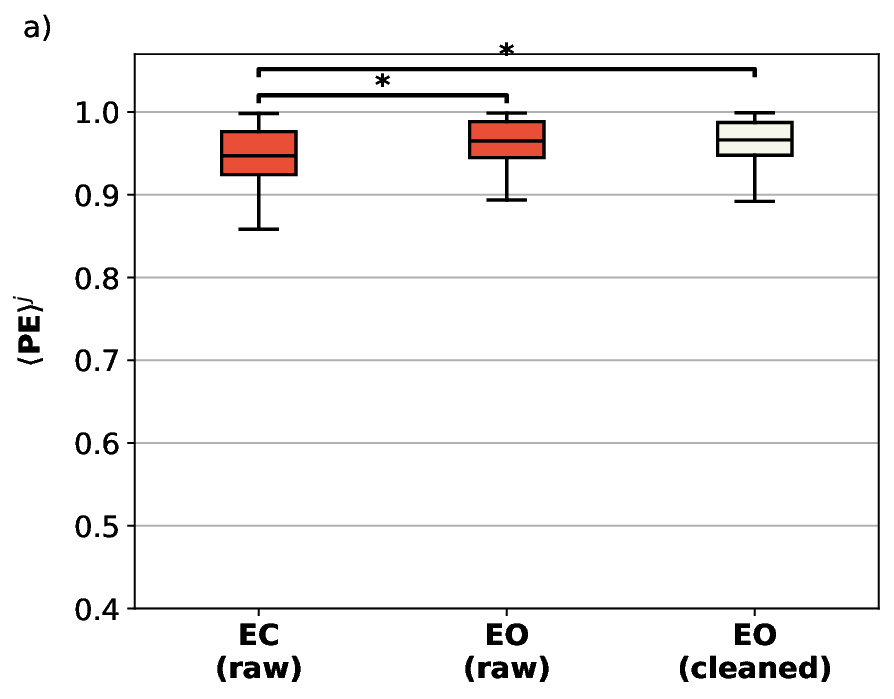}
    \includegraphics[width=0.75\linewidth]{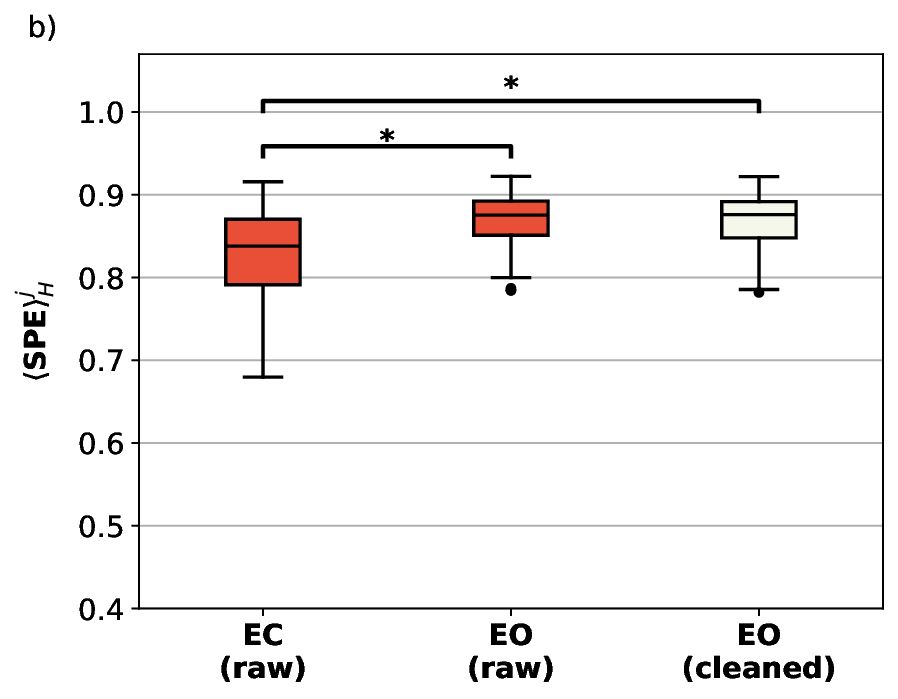}
    \includegraphics[width=0.75\linewidth]{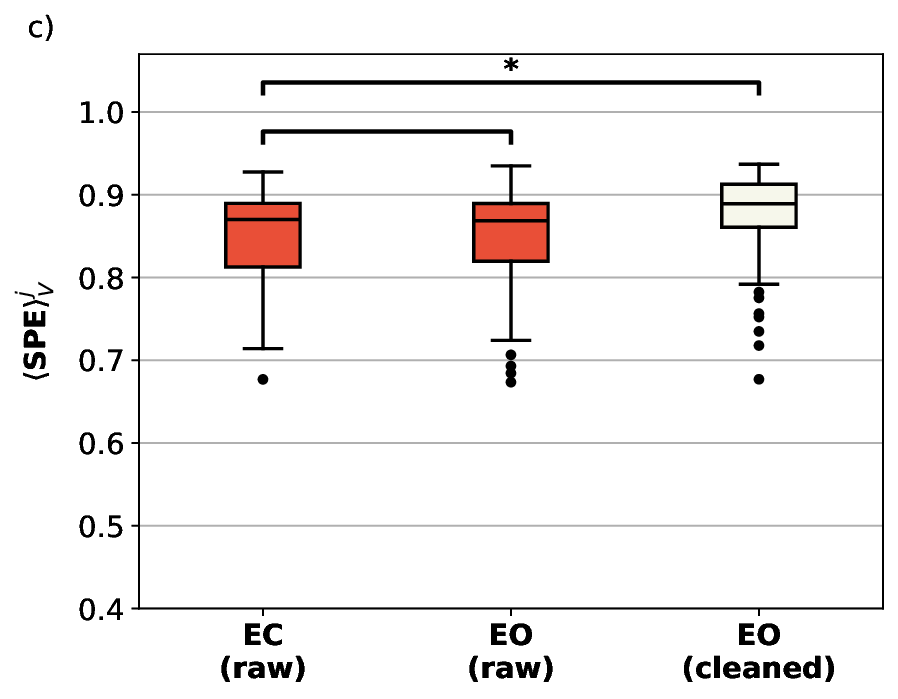}
    \caption{As Fig.~\ref{fig:PE_boxplot} but when 31 channels are analyzed. }
    \label{fig:SPE_montages}
\end{figure}

\begin{figure}[tb] %fig. 8
    \centering
    \includegraphics[width=0.75\linewidth]{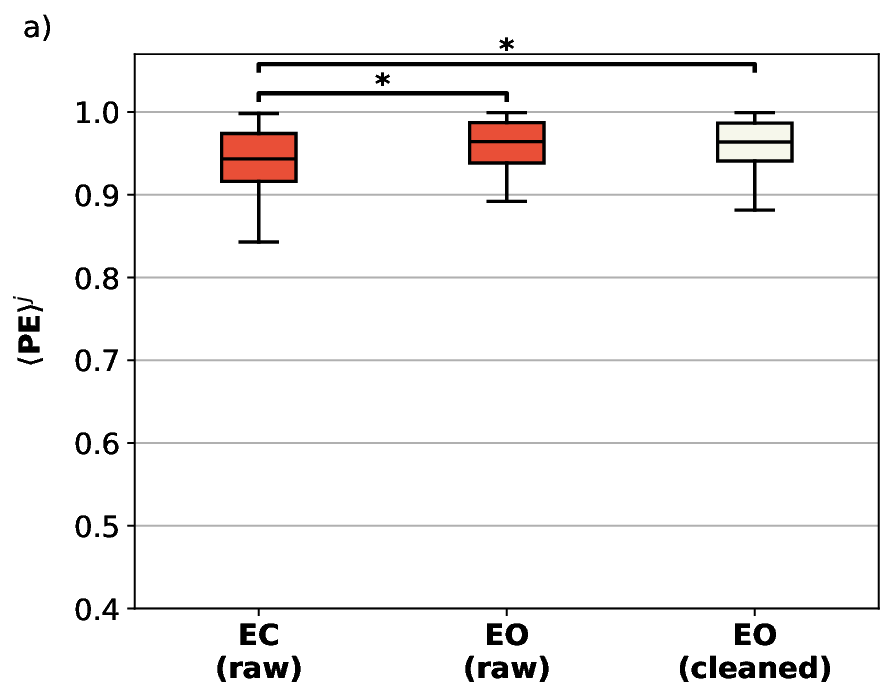}
    \includegraphics[width=0.75\linewidth]{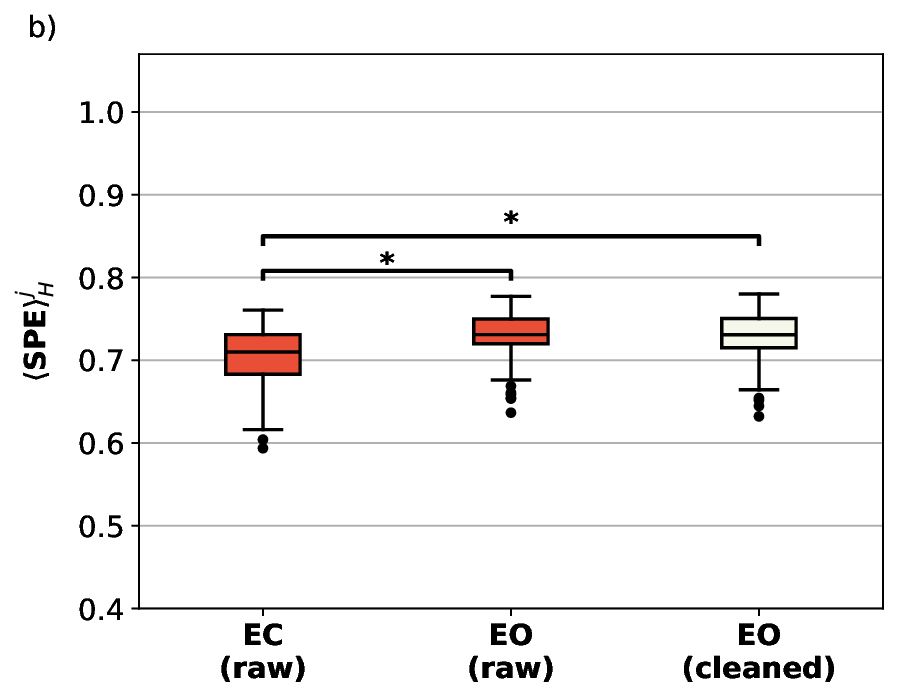}
    \includegraphics[width=0.75\linewidth]{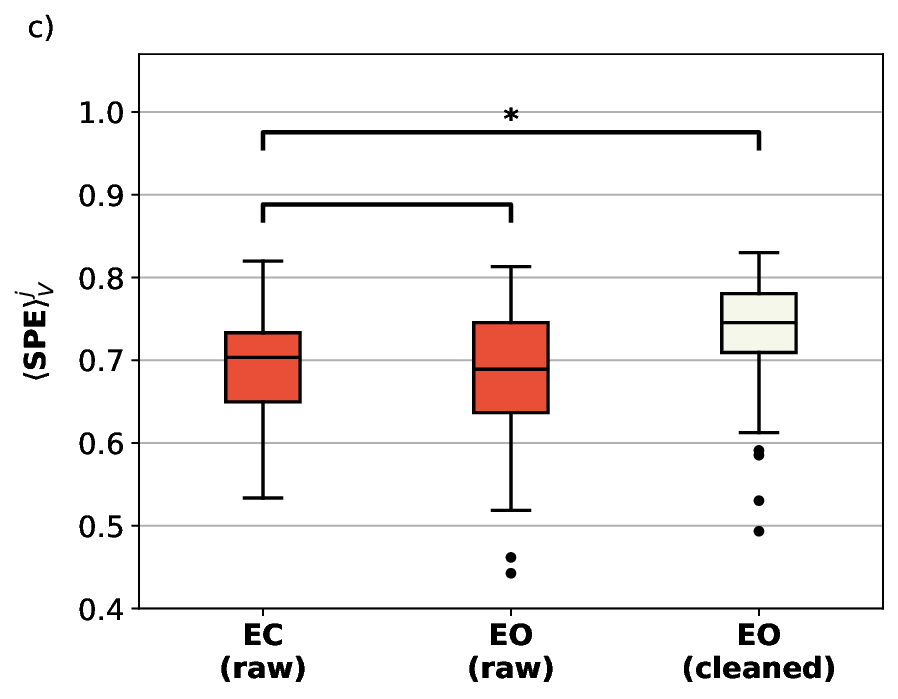}
    \caption{As Fig.~\ref{fig:PE_boxplot} but when 17 channels are analyzed. }
    \label{fig:SPE_montages_17}
\end{figure}

\subsection{Effect of the time interval analyzed}

Because $SPE$ is calculated from spatial OPs defined over a single snapshot of the data values in the EEG channels at a given time, we now analyze how many snapshots are needed, so that the distribution of spatial $PE$ values, $\langle SPE\rangle^{j}$ with $j=1,\dots,109$, reveals differences between the EO and EC states, and we also analyze how long the recording must be for the distribution of temporal $PE$ values, $\langle PE\rangle^{j}$, to capture EO/EC differences. %(we remind the reader that the probabilities of temporal OPs are estimated from data values in the same channel at different times).

{Specifically, we analyze }
%Figure~\ref{fig:pvalues_as_t} shows 
how the p-values obtained when comparing the EO/EC entropy distributions, {as described in Sec.~\ref{sec:pvalues}}, depend on the time interval analyzed. %($\langle PE\rangle^j$, $\langle SPE\rangle^j_H$ and $\langle SPE\rangle^j_V$) to detect significant differences in the EO and EC states, as a function of the length of the recordings analyzed. 

{For this analysis, we divided each time series in 118 non-overleaping segments of $0.5$~s each %($118\times 0.5$~s $=59$~s, and as we explained in Sec.\ref{sec:data}, we disregarded the last second of each recording), 
which contain 80 data points each. In Fig.~\ref{fig:pvalues_as_t} we report the median (as a continuous line) of the distribution of p-values obtained in the different segments, and the spread between first and third quartiles (as shaded areas).}

{Figure~\ref{fig:pvalues_as_t}a shows the p-values obtained for $\langle PE\rangle^{j}$, as a function of the time interval of each segment analyzed. We note that the horizontal axis starts at $0.025$~s, which is the minimum time interval that gives a non-zero value of the entropy: it contains 4 data points %where the first value for each epoch is delayed $0.025s$ since we need at least $4$ data points to obtain the first non-trivial result 
and we need $3$ data points to define the one ordinal pattern, and 2 patterns to obtain a value of the entropy different from zero. }

In Fig.~\ref{fig:pvalues_as_t}a we see that, for the raw data, the median of the distribution of p-values is below the significance level ($10^{-3}$) when we analyze only the first $0.12$~s of each segment, which contains 19 data points that in turn, allow the definition of $17$ ordinal patterns. For the cleaned data, the median p-value decreases even faster and is below the significant level with only 4 ordinal patterns. This is unexpected, because with only 4 ordinal patterns, the distribution of ordinal patterns is under sampled. An examination of their probabilities reveals that in the EC state, patterns A and E tend to be more expressed. %\textcolor{red}{Additionally, we observe that the p-values obtained from filtered data is lower than from raw data), but in log scales the absolute differences are actually small. As in this case the p-value is computed only with $0.5$ seconds of data at most, in such a short time an artifact can have a significant impact.}% 

%\textcolor{blue}{For this analysis, we have divided our time series into $118$ $0.5s$-long epochs, and here we report the median (as continuous lines) of the distribution of p-values obtained from the different epochs, and the spread between first and third quartiles (as shaded areas).
%In Fig.~\ref{fig:pvalues_as_t}a we show the results obtained for $\langle PE\rangle^{j}$, where the first value for each epoch is delayed $0.025s$ since we need at least $4$ data points to obtain the first non-trivial result (we need at least $3$ data points to construct the first symbols, and at least $2$ symbols to obtain a non-zero entropy). For raw data, we can see that the median is already below the significance level around $0.12s$, which allows the construction of $17$ different symbols. For the cleaned data, we are below the significant threshold with only $4$ symbols. This is quite of a surprising result, as in this case we are severely under sampling the underlying distribution of ordinal patterns.} % Nevertheless, since the filtering of the artifacts is done only in the EO states, we are ... }
%In Fig.~\ref{fig:pvalues_as_t}a the analysis starts when the recordings start ($t=0$). We see that all entropies except $\langle SPE\rangle^j_V$ calculated from the raw data (with artifacts) rapidly capture ($<0.2$s) significant differences between EO/EC states. We highlight how artifact removal results in an increase of one order of magnitude in the significance of $SPE_V$ analysis, while for $SPE_H$, it has very little effect.

Figure~\ref{fig:pvalues_as_t}b displays the results obtained for $\langle SPE\rangle^{j}_H$, where we observe that the removal of the artifacts makes almost no difference in the distribution of p-values, and we also note that the evaluation of the spatial entropy in only one time step allows us to obtain a distribution of p-values that is below the significance level ($10^{-3}$). {Because we perform a paired statistical test (EO/EC differences in the recordings of the same subject), we speculate that differences can be detected in ``real-time", from the analysis of the data values recorded at a single time.}

{Figure~\ref{fig:pvalues_as_t}c displays the results obtained for $\langle SPE\rangle^{j}_V$, where significant differences between EO and EC states are only seen in the cleaned data. For this case, significant level is obtained only after 2 time steps, slightly worse than for $\langle SPE\rangle^{j}_H$.}

%\textcolor{blue}{In Fig.~\ref{fig:pvalues_as_t}b we show the results for $\langle PE\rangle^{j}_H$, where we observed not only that the removal of the artifacts makes no different in the distinction between states, but there is a significant difference between states after the evaluation of only one time step. Such result is highly relevant, as it implies that these differences can be captured in a ``real-time" fashion, in the sense that differences in the entropy values can be detected from individual time steps of the data. Finally, in Fig.~\ref{fig:pvalues_as_t}c we show the results for $\langle PE\rangle^{j}_V$, where significant differences between states are only found in the cleaned data. For this case, significant level is attained only after 4 time steps, same as for $\langle PE\rangle^{j}$.}

{To evaluate whether there are transient effects at the start of the recordings, we have compared the results obtained from segments at the beginning or at the end of the recordings, but no significant differences were found.  Except for $\langle SPE\rangle^{j}_V$ from raw data, all quantifiers %perform exceptionally well, detecting 
detect significant differences between EO and EC states from very short time series ($t\lesssim 0.12$~s), with $\langle SPE\rangle^{j}_H$ being able to differentiate the state using only one time step and raw data. }%Careful interpretation is required for the good results of $\langle PE\rangle^{j}$, as the entropy is obtained from a severely under-sampled distribution of ordinal patterns. 

{Regarding the interpretation of these results, one could argue that the artifact removal, which was done only in EO recordings, could alter some time series features, thus allowing a trivial distinction of EO and EC states; however, we believe that this is not the case because in the analysis of $\langle SPE\rangle^{j}_H$ (Fig.~\ref{fig:pvalues_as_t}b), there is almost no effect of artifact removal. }

%\textcolor{blue}{To evaluate whether there are transient effects at the start of the recordings, we have evaluated separately epochs from early and latter stages of the experiment, but no differences were observed for any of the three measures.  Except for $\langle SPE\rangle^{j}_V$ from raw data, all quantifiers perform exceptionally well, detecting significant differences between states from very short time series ($t\lesssim 0.12$), with $\langle SPE\rangle^{j}_H$ being able to do so for single time steps, and even from raw data. Careful interpretation is required for the good results of $\langle PE\rangle^{j}$, as the entropy is obtained from a severely under-sampled distribution of ordinal patterns. Because as the removing of the artifacts is done only in EO states, while EC states are left raw, might happen that the filtering is removing not only the artifacts but additional features of the time series, which trivializes the distinction between states. This could also apply to $\langle SPE\rangle^{j}_V$, but at the same time strengthens the robustness of $\langle SPE\rangle^{j}_H$, as no effect is observed in this case. }

\subsection{Effect of the number of channels analyzed}

We conclude by analyzing how robust the permutation entropy approach is with respect to the number of channels analyzed. %For this, we consider two subsets of the electrodes, both shown in Fig.~\ref{fig:montage}. In 
Figures~\ref{fig:SPE_montages} and \ref{fig:SPE_montages_17} display the entropy distributions (as in Fig.~\ref{fig:PE_boxplot}) but when only 31 or 17 electrodes are analyzed, respectively. Similar results are obtained in all cases. %, only a significant decrease in the p-values is observed for the SPE obtained from horizontal symbols in the 17-electrode montage (Fig.~\ref{fig:SPE_montages}c). 
In the spatial approach (panels b and c), because the number of spatial OPs available to estimate the probabilities is low or very low (see Table~\ref{tab:number_of_patterns}), the under-sampling of the distribution of spatial OPs lowers the $SPE$ values; however, the effect is the same in the EO and EC states and therefore, the differences between the entropy values remain significant.
{In the temporal approach, differences are captured in single channels located in the posterior (over the visual cortex) and central regions (as shown in Fig.~\ref{fig:topomaps_PE}), as it was already reported in Quintero-Quiroz et al.\cite{QuinteroQuiroz2018}.}

\section{Conclusions and discussion}

{In this work we studied the robustness of permutation entropy analysis to differentiate closed-eyes (EC) and open-eyes (EO) states, comparing the temporal ($PE$) and spatial ($SPE$) approaches. Previous studies have shown that EC states tend to have lower entropy values than EO states \citep{QuinteroQuiroz2018,Boaretto2023,Gancio2024}. Here we analyzed the robustness of $PE$ and $SPE$ quantifiers, in relation to the presence of blinking artifacts in the EO state, the length of the EEG recordings, and the number of EEG channels analyzed. }%in electroencephalogram (EEG) signals, applying such quantifiers to the problem of distinguishing two different brain resting states: one with the eyes open (EO), and the other with the eyes closed (EC) and require complex pre-processing in order to be removed. 

%Previous works \citep{QuinteroQuiroz2018,Boaretto2023} have shown that these quantifiers are able to distinguish these states, as EC states present lower entropy values than EO states.
{Blinking artifacts are ubiquitous in EEG recordings %while the subjects keep their eyes open, require complex pre-processing in order to be removed, and hugely affect many commonly used statistical features. In contrast, 
and we have shown that, for the data analyzed here, $PE$ is barely affected by these artifacts. We speculate that this robustness is due to the fact that the procedure for artifact removal preserves {the nonlinear statistical properties of the signal; hence, the obtained distribution of ordinal patterns is very similar, and} artifact removal has little or no effect on $PE$ values. Therefore, our results suggest that $PE$ can be used {directly on the raw} EEG signals, saving pre-processing time.} %, as they preserve the small scale structure of the signal. This robustness is based in one of the core features of $PE$, as its calculation is independent of the actual values of the time series, taking into consideration only the relative ordering of consecutive data points. Such robustness does not only exhibit the reliability of $PE$ as a robust feature of the brain's dynamics, but also its ability to work on the raw data yielding accurate results, which could be used in on-line applications where fast analysis is required, and skipping the pre-processing saves valuable time. 

We have also shown that $SPE$ is robust to blinking artifacts, with the orientation of the spatial ordinal patterns playing a key role. {Since a blinking artifact forms a gradient with anterior-posterior orientation along the pre-frontal and frontal channels, the spatial ordinal patterns defining $SPE_V$ will be affected. In particular, the pattern 1-2-3 (labeled "A" in Fig. 1d) will be over-expressed, and the pattern 3-2-1 ("F") will be under-expressed.} Because blinking artifacts only occur while subjects have their eyes open, these over-expressed and under-expressed ordinal patterns lower the entropy of the EO states, making it similar to that of EC states, hindering the ability of $SPE_V$ to discriminate between the two. In the case of $SPE_H$, which is calculated from the probabilities of ordinal patterns that have lateral-medial orientation, and therefore are perpendicular to the gradient induced by the artifacts, the $\langle SPE\rangle^{j,EO}_H$ values are almost not affected by them. This keeps $\langle SPE\rangle^{j,EO}_H$ unaltered and significantly higher than $\langle SPE\rangle^{j,EC}_H$. These results are consistent with our earlier work \cite{Gancio2024}, where we found that $\langle SPE\rangle_V$ had poor performance when applied to raw signals, but improved after the signals were filtered to analyze only the $\alpha$-band ($8-12$~Hz), which is not affected by blinking artifacts.

%We have also shown that $SPE$ is robust to blinking artifacts, but the orientation of the spatial symbols plays a key role in this ability. Because during these artifacts an anterior-posterior gradient is created in the data, the spatial patterns in this orientation (which produce $SPE_V$) are strongly conditioned by this gradient, over-expressing the 1-2-3 symbol (constantly increasing values), while under expression the 3-2-1 symbol (constantly decreasing values). Because blinking artifacts only occur while subjects have their eyes open, this conditioning of the symbols lowers the entropy of the EO states (which is usually higher than the one for EC states), making it similar to the one for EC states, hindering the ability of $SPE_V$ to discriminate between these states. In the case of $SPE_H$, where the patterns have lateral-medial orientation and therefore are perpendicular to the gradient induced by the artifacts and the values obtained for $\langle SPE\rangle^{j,EO}_H$ are not affected, maintaining the ability of this quantifier to differentiate between states. This explains the results obtained in \cite{Gancio2024}, where $\langle SPE\rangle_V$ performed poorly in the classification between EO and EC raw signals, but once the signals were filtered to analyze only the $\alpha$-band ($8-12Hz$). This filtering also removes the blinking artifacts, restoring the capacity of $\langle SPE\rangle_V$ to distinguish between states. 

{We also analyzed how these quantifiers are affected by the amount of data provided, either by restricting the length of the time series, or the number of channels analyzed. Surprisingly, all the entropy quantifiers (except $\langle SPE\rangle_V$ from the raw data, which is affected by the artifacts) are able to capture differences between the brain states from extremely short time series ($\lesssim0.12$~s). $SPE_H$ %outperforms $PE$ because it 
captures significant differences in single snapshots of the data values in the EEG channels, however, $PE$ also detects differences surprisingly fast. }

{However, regarding the performance of $PE$ on cleaned data, this result has to be critically considered, because the filtering is applied only to EO recordings, and it could be modifying the time series in a way that facilitates the distinction between EO and EC states. To further clarify this point, other artifact removal techniques \cite{Barban2021} should be used to confirm that $PE$ performance is unrelated to the specific procedure for artifact removal.} %Overall, we have found that $PE$ and $SPE$ can process information similarly for time series of similar length. ESTO HABRIA QUE CUANTIFICAR

%Finally, we have also shown how these quantifiers are affected by the amount of data provided, by restricting the length of the time series, and the number of channels analyzed. Surprisingly, all the measures considered (except $\langle SPE\rangle_V$ from the raw data, which is affected by the artifacts) are able to capture differences between the brain states from extremely short time series ($\lesssim0.12s$). Although $SPE_H$ outperforms $PE$, since the former captures significant differences in single snapshots of the brain, $PE$ performs surprisingly fast. However, for the case of $PE$ from cleaned data, this result has to be carefully considered, as filtering could be removing more than just artifacts. In this sense, other artifact removal techniques\cite{Barban2021} could be applied to confirm that this fast performance is not related to the specifics of the ICA approach. Overall, we have found that $PE$ and $SPE$ can process information similarly for time series of similar length. 

When restricting the number of channels analyzed, from 64 to 31 or 17 channels, we found that all entropy quantifiers {still} detect significant differences between the EO and EC states. %perform as well, even when analyzing less than one third of the original channels. 
For $PE$ this could be expected, because we sub-sampled the channels keeping electrodes in all brain regions. As shown in Fig.~\ref{fig:topomaps_PE}, as long as we keep channels records in the posterior region, where the visual cortex is located, $PE$ can detect significant differences between states. %\textcolor{red}{(HOW MANY CHANNELS WE NEED TO KEEP IN THE VISUAL CORTEX REGION? IS ONE CHANNEL IN THE VISUAL CORTEX ENOUGH? THIS IS NEW INFO, SO THIS SHOULD BE FIRST DISCUSSED IN SECTION IVC).} \textcolor{blue}{De acuerdo con Fig.3, toda la región es significativa, en la versión inicial de esta fig se incluia un mapa topografico con los p-valores, pero lo quitamos. Este resultado ya se reporta en el paper de Quintero-Quiroz, pero lo podemos mencionar en la Sec. IVC.}

Regarding $SPE$, it also detects significant differences when the number of channels analyzed are only 31 or 17. The underlying reason is not yet understood and further analyses are needed to clarify this point, but we speculate it can be related to the fact that {undersampling the channels introduces a negative bias to the entropy, that for a given subject, in principle, will be the same for both the EO and EC states. Since we are interested only in the \emph{difference} in the $SPE$ values, the two biases should approximately cancel out, leaving the significance levels of the differences almost unaltered.}%when analyzing very short time intervals (in which very few patterns can be defined and therefore, the six ordinal probabilities are strongly under-sampled), significant differences between EC and EO states were detected by the entropy quantifies.

%When restricting the number of channels, we found that all quantifiers perform as well, even when analyzing less than one third of the original channels. For $PE$ this was expected, as in this sub-sampling of channels we kept electrodes from every region of the brain. As shown in Fig.~\ref{fig:topomaps_PE}, as long as we keep channels that sample the posterior region, where the visual cortex is located, we can detect differences between states. For $SPE$, which also preserves similar performance as channels are removed, the results must be analyzed with more care, as in this case the available sampling of the different patterns used for the estimation of probabilities is extremely compromised. This compromise is already made when considering the analysis of different orientations, as restricting to only rows or columns yields only 45 and 44 available patterns to compute the probabilities of 6 symbols. This under-sampling is more severe as the channels are removed. 

Taken together, our analysis shows that $PE$ and $SPE$ quantifies can be used even when the distribution of temporal or spatial ordinal patterns is under-sampled. However, our results also leave a number of important open questions for future work. First, it will be interesting to consider a very small number of electrodes, localized in a small region of the brain, to uncover which region is the most informative. This study could have practical application for small portable devices, brain computer interfacing (BCI) or medical sensors that monitor small regions of the scalp. Second, it will be interesting to consider which is the optimal trade-off, regarding the number of electrodes and the duration of the time series segment analyzed. We have found here that with a single ``snapshot'' of 64 channels, $SPE_H$, can detect significant differences; however, which is the minimum number of channels needed? Third, a natural next step is to quantify the classification performance and analyze how it depends on the number of channels analyzed, the duration of the time interval analyzed, and the length of the ordinal patterns (here we have used $L=3$ data points recorded in the same channel at consecutive times, or at the same time in different channels, but one could increase $L$ and combine the temporal and spatial approaches). {Fourth, spatial ordinal patterns also enable the investigation of the two-dimensional spatial organization of the cerebral cortex, either through the analysis of genuine 2D patterns\citep{Bandt_2023,weiss2025non} or through representations based on space-filling Hilbert curves\cite{bariviera2025texture}. However, this approach remains relatively underexplored and therefore constitutes a promising direction for future research.} Last but not least, it will also be interesting to compare the permutation entropy approach with extensions of permutation entropy, as weighted-permutation entropy\citep{fadlallah2013weighted} or amplitude-aware permutation entropy\citep{azami2016amplitude}, and with other entropy-based quantifiers, as the OP variability\citep{politi2017quantifying,tyloo2025including}, or ordinal transitions\citep{olivares2020,leyva2023}.

\section{Acknowledgments}
This work was partially funded by the Agencia Estatal de Investigacion (PID2024-160573NB-I00). J. G. also acknowledges support of Agencia de Gestió d’Ajuts Universitaris i de Recerca (FI AGAUR 2023 FI-1 00034), and GT acknowledges support from the Serra Húnter Programme (Generalitat de Catalunya). % and C. M, of the Institució Catalana de Recerca i Estudis Avançats (ICREA Academia).

\section{Data availability}
The dataset analyzed for this study can be found in the PhysioNet: {\url{https://physionet.org/content/eegmmidb/1.0.0/}}

\section*{References}
\bibliography{Bibliography_Introduction}

\end{document}